\documentclass[12pt,a4paper]{article}

\usepackage{ifthen} 
\newboolean{pdflatex}
\setboolean{pdflatex}{true} 
\usepackage{rotating}
\usepackage{afterpage}

\newboolean{articletitles}
\setboolean{articletitles}{true} 

\newboolean{uprightparticles}
\setboolean{uprightparticles}{false} 

\def\paperauthors{LHCb collaboration} 
\def\paperasciititle{Measurement of $b$-hadron fractions in 13 TeV pp collisions} 
\def\papertitle{Measurement of $b$-hadron fractions in $13\tev$ $pp$ collisions} 
\def\paperkeywords{{High Energy Physics}, {LHCb}} 
\def\papercopyright{\the\year\ CERN for the benefit of the LHCb collaboration} 
\def\paperlicence{CC-BY-4.0 licence}
\def\paperlicenceurl{https://creativecommons.org/licenses/by/4.0/}

\usepackage[top=1in, bottom=1.25in, left=1in, right=1in]{geometry}

\usepackage{microtype}
\usepackage{lineno}  
\usepackage{xspace} 
\usepackage{caption} 

\usepackage{graphicx}  
\usepackage{color}
\usepackage{colortbl}
\graphicspath{{./figs/}} 
\DeclareGraphicsExtensions{.pdf,.PDF,png,.PNG}

\usepackage{amsmath} 
\usepackage{amssymb}
\usepackage{amsfonts}
\usepackage{upgreek} 

\newcommand*\patchAmsMathEnvironmentForLineno[1]{%
\expandafter\let\csname old#1\expandafter\endcsname\csname #1\endcsname
\expandafter\let\csname oldend#1\expandafter\endcsname\csname
end#1\endcsname
 \renewenvironment{#1}%
   {\linenomath\csname old#1\endcsname}%
   {\csname oldend#1\endcsname\endlinenomath}%
}
\newcommand*\patchBothAmsMathEnvironmentsForLineno[1]{%
  \patchAmsMathEnvironmentForLineno{#1}%
  \patchAmsMathEnvironmentForLineno{#1*}%
}
\AtBeginDocument{%
\patchBothAmsMathEnvironmentsForLineno{equation}%
\patchBothAmsMathEnvironmentsForLineno{align}%
\patchBothAmsMathEnvironmentsForLineno{flalign}%
\patchBothAmsMathEnvironmentsForLineno{alignat}%
\patchBothAmsMathEnvironmentsForLineno{gather}%
\patchBothAmsMathEnvironmentsForLineno{multline}%
\patchBothAmsMathEnvironmentsForLineno{eqnarray}%
}

\usepackage{hyperxmp}

\usepackage[pdftex,
            pdfauthor={\paperauthors},
            pdftitle={\paperasciititle},
            pdfkeywords={\paperkeywords},
            pdfcopyright={Copyright (C) \papercopyright},
            pdflicenseurl={\paperlicenceurl}]{hyperref}

\usepackage[colorinlistoftodos,textsize=scriptsize]{todonotes}

\usepackage[all]{hypcap} 

\usepackage{xspace} 
\usepackage{upgreek}

\newcommand{\offsetoverline}[2][0.1em]{\kern #1\overline{\kern -#1 #2}}%

\def\lhcb   {\mbox{LHCb}\xspace}

\def\MagUp {\mbox{\em Mag\kern -0.05em Up}\xspace}

\ifthenelse{\boolean{uprightparticles}}%
{

 \def\Pmu         {\ensuremath{\upmu}\xspace}                 
 \def\Pnu         {\ensuremath{\upnu}\xspace}                 
                  
 \def\Ppi         {\ensuremath{\uppi}\xspace}

 \def\Ppsi        {\ensuremath{\uppsi}\xspace}

 \def\PDelta      {\ensuremath{\Delta}\xspace}                 
 \def\PXi         {\ensuremath{\Xi}\xspace}                 
 \def\PLambda     {\ensuremath{\Lambda}\xspace}                 
 \def\PSigma      {\ensuremath{\Sigma}\xspace}                 
 \def\POmega      {\ensuremath{\Omega}\xspace}                 
 \def\PUpsilon    {\ensuremath{\Upsilon}\xspace}

 \def\PB      {\ensuremath{\mathrm{B}}\xspace}                 
                  
 \def\PD      {\ensuremath{\mathrm{D}}\xspace}

 \def\PJ      {\ensuremath{\mathrm{J}}\xspace}                 
 \def\PK      {\ensuremath{\mathrm{K}}\xspace}

 \def\Pb      {\ensuremath{\mathrm{b}}\xspace}                 
 \def\Pc      {\ensuremath{\mathrm{c}}\xspace}

 \def\Pi      {\ensuremath{\mathrm{i}}\xspace}

 \def\Ps      {\ensuremath{\mathrm{s}}\xspace}

}
{

 \def\Pmu         {\ensuremath{\mu}\xspace}                 
 \def\Pnu         {\ensuremath{\nu}\xspace}                 
                  
 \def\Ppi         {\ensuremath{\pi}\xspace}

 \def\Ppsi        {\ensuremath{\psi}\xspace}                 
                  
 \mathchardef\PDelta="7101
 \mathchardef\PXi="7104
 \mathchardef\PLambda="7103
 \mathchardef\PSigma="7106
 \mathchardef\POmega="710A
 \mathchardef\PUpsilon="7107
                  
 \def\PB      {\ensuremath{B}\xspace}                 
                  
 \def\PD      {\ensuremath{D}\xspace}

 \def\PJ      {\ensuremath{J}\xspace}                 
 \def\PK      {\ensuremath{K}\xspace}

 \def\Pb      {\ensuremath{b}\xspace}                 
 \def\Pc      {\ensuremath{c}\xspace}

 \def\Pi      {\ensuremath{i}\xspace}

 \def\Ps      {\ensuremath{s}\xspace}

}

\makeatletter
\ifcase \@ptsize \relax
  \newcommand{\miniscule}{\@setfontsize\miniscule{4}{5}}
\or
  \newcommand{\miniscule}{\@setfontsize\miniscule{5}{6}}
\or
  \newcommand{\miniscule}{\@setfontsize\miniscule{5}{6}}
\fi
\makeatother

\DeclareRobustCommand{\optbar}[1]{\shortstack{{\miniscule (\rule[.5ex]{1.25em}{.18mm})}
  \\ [-.7ex] $#1$}}

\def\mup        {{\ensuremath{\Pmu^+}}\xspace}
\def\mun        {{\ensuremath{\Pmu^-}}\xspace} 

\def\neub       {{\ensuremath{\overline{\Pnu}}}\xspace}

\def\neumb      {{\ensuremath{\neub_\mu}}\xspace}

\def\squark    {{\ensuremath{\Ps}}\xspace}

\def\cquark    {{\ensuremath{\Pc}}\xspace}

\def\bquark    {{\ensuremath{\Pb}}\xspace}

\def\pion   {{\ensuremath{\Ppi}}\xspace}

\def\pip    {{\ensuremath{\pion^+}}\xspace}

\def\kaon    {{\ensuremath{\PK}}\xspace}
  \def\Kbar    {{\kern 0.2em\overline{\kern -0.2em \PK}{}}\xspace}
\def\Kb      {{\ensuremath{\Kbar}}\xspace}
\def\KorKbar {\kern 0.18em\optbar{\kern -0.18em K}{}\xspace}
\def\Kz      {{\ensuremath{\kaon^0}}\xspace}

\def\Kp      {{\ensuremath{\kaon^+}}\xspace}
\def\Km      {{\ensuremath{\kaon^-}}\xspace}

  \def\Dbar    {{\kern 0.2em\overline{\kern -0.2em \PD}{}}\xspace}
\def\D       {{\ensuremath{\PD}}\xspace}
\def\Db      {{\ensuremath{\Dbar}}\xspace}
\def\DorDbar {\kern 0.18em\optbar{\kern -0.18em D}{}\xspace}
\def\Dz      {{\ensuremath{\D^0}}\xspace}

\def\Dp      {{\ensuremath{\D^+}}\xspace}

\def\Dstar   {{\ensuremath{\D^*}}\xspace}

\def\Ds      {{\ensuremath{\D^+_\squark}}\xspace}

\def\B       {{\ensuremath{\PB}}\xspace}
\def\Bbar    {{\ensuremath{\kern 0.18em\overline{\kern -0.18em \PB}{}}}\xspace}
\def\Bb      {{\ensuremath{\Bbar}}\xspace}
\def\BorBbar    {\kern 0.18em\optbar{\kern -0.18em B}{}\xspace}

\def\Bzb     {{\ensuremath{\Bbar{}^0}}\xspace}

\def\Bub     {{\ensuremath{\B^-}}\xspace}

\def\Bm      {{\ensuremath{\Bub}}\xspace}

\def\Bsb     {{\ensuremath{\Bbar{}^0_\squark}}\xspace}

\def\jpsi     {{\ensuremath{{\PJ\mskip -3mu/\mskip -2mu\Ppsi\mskip 2mu}}}\xspace}

\def\Y#1S{\ensuremath{\PUpsilon{(#1S)}}\xspace}

\def\Lz          {{\ensuremath{\PLambda}}\xspace}

\def\LorLbar     {\kern 0.18em\optbar{\kern -0.18em \PLambda}{}\xspace}

\def\Lc          {{\ensuremath{\Lz^+_\cquark}}\xspace}

\def\Lb           {{\ensuremath{\Lz^0_\bquark}}\xspace}

\def\to                 {\ensuremath{\rightarrow}\xspace}

\def\AT#1     {\ensuremath{A_{\mathrm{T}}^{#1}}\xspace}           

\def\C#1      {\ensuremath{\mathcal{C}_{#1}}\xspace}                       
\def\Cp#1     {\ensuremath{\mathcal{C}_{#1}^{'}}\xspace}                    
\def\Ceff#1   {\ensuremath{\mathcal{C}_{#1}^{\mathrm{(eff)}}}\xspace}        
\def\Cpeff#1  {\ensuremath{\mathcal{C}_{#1}^{'\mathrm{(eff)}}}\xspace}       
\def\Ope#1    {\ensuremath{\mathcal{O}_{#1}}\xspace}                       
\def\Opep#1   {\ensuremath{\mathcal{O}_{#1}^{'}}\xspace}                    

\newcommand{\aunit}[1]{\ensuremath{\text{\,#1}}}       

\newcommand{\tev}{\aunit{Te\kern -0.1em V}\xspace}
\newcommand{\gev}{\aunit{Ge\kern -0.1em V}\xspace}
\newcommand{\mev}{\aunit{Me\kern -0.1em V}\xspace}
\newcommand{\kev}{\aunit{ke\kern -0.1em V}\xspace}
\newcommand{\ev}{\aunit{e\kern -0.1em V}\xspace}
\newcommand{\mevc}{\ensuremath{\aunit{Me\kern -0.1em V\!/}c}\xspace}
\newcommand{\gevc}{\ensuremath{\aunit{Ge\kern -0.1em V\!/}c}\xspace}
\newcommand{\mevcc}{\ensuremath{\aunit{Me\kern -0.1em V\!/}c^2}\xspace}
\newcommand{\gevcc}{\ensuremath{\aunit{Ge\kern -0.1em V\!/}c^2}\xspace}

\def\fb   {\ensuremath{\aunit{fb}}\xspace}
\def\invfb   {\ensuremath{\fb^{-1}}\xspace}

\def\fs   {\aunit{fs}}

\newcommand{\chisq}{\ensuremath{\chi^2}\xspace}

\newcommand{\chisqip}{\ensuremath{\chi^2_{\text{IP}}}\xspace}

\def\gsim{{~\raise.15em\hbox{$>$}\kern-.85em
          \lower.35em\hbox{$\sim$}~}\xspace}
\def\lsim{{~\raise.15em\hbox{$<$}\kern-.85em
          \lower.35em\hbox{$\sim$}~}\xspace}

\def\pt         {\ensuremath{p_{\mathrm{T}}}\xspace}

\def\evtgen     {\mbox{\textsc{EvtGen}}\xspace}

\def\geant      {\mbox{\textsc{Geant4}}\xspace}

\def\photos     {\mbox{\textsc{Photos}}\xspace}

\def\pythia     {\mbox{\textsc{Pythia}}\xspace}

\xspace

\def\tell1  {TELL1\xspace}
\def\ukl1   {UKL1\xspace}

\usepackage{cite} 
\usepackage{mciteplus}

\usepackage{longtable} 

\def\fs{\ensuremath{f_s\xspace}}
\def\fd{\ensuremath{f_d\xspace}}
\def\fu{\ensuremath{f_u\xspace}}
\def\fp{\ensuremath{f_+\xspace}}
\def\fz{\ensuremath{f_0\xspace}}

\def\rbs{f_s/(f_u+f_d)}

\def\munu{\mu^-\overline{\nu}_{\mu}}
\DeclareRobustCommand{\myoptbar}[1]{\shortstack{{\miniscule (\rule[.5ex]{0.85em}{.18mm})}
  \\ [-.7ex] $#1$}}
\def\porpbar    {\kern 0.02em\myoptbar{\kern -0.02em p}{}\xspace}
\begin{document}

\renewcommand{\thefootnote}{\fnsymbol{footnote}}
\setcounter{footnote}{1}


\begin{titlepage}
\pagenumbering{roman}

\vspace*{-1.5cm}
\centerline{\large EUROPEAN ORGANIZATION FOR NUCLEAR RESEARCH (CERN)}
\vspace*{1.5cm}
\noindent
\begin{tabular*}{\linewidth}{lc@{\extracolsep{\fill}}r@{\extracolsep{0pt}}}
\ifthenelse{\boolean{pdflatex}}
{\vspace*{-1.5cm}\mbox{\!\!\!\includegraphics[width=.14\textwidth]{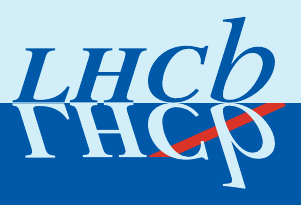}} & &}%
{\vspace*{-1.2cm}\mbox{\!\!\!\includegraphics[width=.12\textwidth]{lhcb-logo.eps}} & &}%
\\
 & & CERN-EP-2019-016 \\  
 & & LHCb-PAPER-2018-050 \\  
 & & February 18, 2019 \\ 
 & & \\
\end{tabular*}

\vspace*{4.0cm}

{\normalfont\bfseries\boldmath\huge
\begin{center}
  \papertitle 
\end{center}
}

\vspace*{2.0cm}

\begin{center}
\paperauthors\footnote{Authors are listed at the end of this paper.}
\end{center}

\vspace{\fill}

\begin{abstract}
  \noindent
The production fractions of  \Bsb and \Lb hadrons, normalized to the sum of \Bm and \Bzb fractions, are measured in $13\tev$ $pp$ collisions using data collected by the LHCb experiment, corresponding to an integrated luminosity of 1.67\invfb.  These ratios, averaged over the $b$-hadron transverse momenta from 4 to 25 GeV and pseudorapidity from 2 to 5, are $0.122  \pm 0.006$ for \Bsb,  and $0.259 \pm 0.018$ for \Lb, where the uncertainties arise from both statistical and systematic sources. The \Lb ratio depends strongly on transverse momentum, while the \Bsb ratio shows a mild dependence. Neither ratio shows variations with pseudorapidity. The measurements are made using semileptonic decays to minimize theoretical uncertainties. In addition, the ratio of \Dp to \Dz mesons produced in the sum of \Bzb and \Bm semileptonic decays is determined as $0.359\pm0.006\pm 0.009$, where the uncertainties are statistical and systematic.

\end{abstract}

\vspace*{2.0cm}

\begin{center}
To be published in Physical Review D Rapid Communications
\end{center}

\vspace{\fill}

{\footnotesize 
\centerline{\copyright~\papercopyright. \href{\paperlicenceurl}{\paperlicence}.}}
\vspace*{2mm}

\end{titlepage}


\newpage
\setcounter{page}{2}
\mbox{~}
%
%
%
%

\cleardoublepage


\renewcommand{\thefootnote}{\arabic{footnote}}
\setcounter{footnote}{0}



\pagestyle{plain} 
\setcounter{page}{1}
\pagenumbering{arabic}


%

Knowledge of the fragmentation fractions of \Bsb ($f_s$) and \Lb ($f_{\Lb}$) hadrons is essential for determining absolute branching fractions (${\cal{B}}$) of decays of these hadrons at the LHC, allowing measurements, for example, of ${\cal{B}}(\Bsb\to\mu^+\mu^-)$ \cite{CMS:2014xfa} and the future evaluation of $|V_{cb}|$ from $\Lb\to\Lc\mu^-\overline{\nu}_{\mu}$ decays \cite{Aaij:2017svr}.\footnote{Mention of a particular decay mode implies the use of the charge-conjugate one as well.} Once these fractions are determined, measurements of absolute branching fractions of \Bm and \Bzb mesons performed at $e^+e^-$ colliders operating at the $\PUpsilon(4S)$ resonance can be used to determine the \Bsb and \Lb branching fractions \cite{Tanabashi:2018oca}. 

In this Letter we measure the ratios  $f_s/(f_u+f_d)$ and $f_{\Lb}/(f_u+f_d)$, where the denominator is the sum of \Bm and \Bzb contributions, in the LHCb acceptance of pseudorapidity $2<\eta <5$ and transverse momentum $4< \pt<25\gev$,\footnote{We use natural units where $c=\hbar=1$.} in 13\tev $pp$ collisions. These ratios can depend on \pt and $\eta$; therefore, we perform the analysis using two-dimensional binning. 

Much of the analysis method adopted in this study is an evolution of our previous \mbox{$b$-hadron} fraction measurements for 7\tev $pp$ collisions\cite{Aaij:2011jp}.  We use the inclusive semileptonic decays $H_b\to H_c X\mu^-\overline{\nu}_{\mu}$, where $H_b$ indicates a $b$ hadron, $H_c$ a charm hadron, and $X$  possible additional particles. Each of the different $H_c$ plus muon final states can
originate from the decay of different $b$ hadrons. Semileptonic decays of $\Bzb$ mesons usually result in a mixture of $D^0$ and $D^+$ mesons, while $B^-$ mesons decay predominantly into $D^0$ mesons with a smaller admixture of $D^+$ mesons. Both include a tiny component of $\Ds \Kb$ meson pairs. Similarly, $\Bsb$ mesons decay predominantly into $D_s^+$ mesons, but can also decay into $D^0K^+$ and $D^+\Kz$ meson pairs; this is expected if the 
$\Bsb$ meson decays into an excited $D_s^+$ state that is heavy enough to decay into a $DK$ pair. We measure this contribution using $D^0K^+X\mu^-\overline{\nu}_{\mu}$ events.  Finally, $\Lb$ baryons decay semileptonically mostly into $\Lc$ final states, but can also decay into $D^0p$ and $D^+n$ pairs. We ignore the contributions of $b\to u$ decays that comprise  approximately 1\% of semileptonic $b$-hadron decays, and contribute almost equally to all $b$-hadron species. The detailed equations relating these yields to the final results are given in Ref.~\cite{Aaij:2011jp} and in the Supplemental material. 

The theoretical basis for this measurement is the near equality of semileptonic widths, $\Gamma_{\rm SL}$, for all $b$-hadron species \cite{Bigi:2011gf} whose differences are predicted to precisions of about 1\%. The values we use for the individual $H_b$ semileptonic branching fractions (${\cal{B}}_{\rm SL}$) are listed in Table~\ref{tab:slpar}. The $H_c$ decay modes used and their branching fractions are given in Table~\ref{tab:charmBF}.

The ratio of \Dp to \Dz meson production in the sum of semileptonic \Bzb and \Bm decays, 
${\fp}/{\fz}$,  is used to check the analysis method. This result can be related to models of the hadronic final states in $\Bm$ and $\Bzb$ semileptonic decays \cite{Rudolph:2018rzl}.

\label{sec:Detector}
The data sample corresponds to 1.67\invfb of integrated luminosity obtained with the LHCb detector in 13\tev $pp$ collisions during 2016. The LHCb 
detector~\cite{Alves:2008zz,LHCb-DP-2014-002} is a
single-arm forward spectrometer covering the pseudorapidity range $2 < \eta < 5$, designed for
the study of particles containing \bquark\ or \cquark\ quarks. The detector elements that are particularly
relevant to this analysis are: a silicon-strip vertex detector surrounding the $pp$ interaction
region that allows \cquark\ and \bquark\ hadrons to be identified from their characteristically long
flight distance from the primary vertex (PV); a tracking system that provides a measurement of the momentum, $p$, of charged
particles, two ring-imaging Cherenkov detectors that are able to discriminate between
different species of charged hadrons, and a muon detection system.

The online event selection is performed by a trigger~\cite{LHCb-DP-2012-004} 
which consists of a hardware stage, based on information from the calorimeter and muon
systems, followed by a software stage, which applies a full event
reconstruction.
At the hardware trigger stage, events are required to have a muon with large \pt or a
  hadron, photon or electron with high transverse energy in the calorimeters. For hadrons,
  the transverse energy threshold is 3.5\gev.
  The software trigger requires a two-, three- or four-track
  secondary vertex with a significant displacement from any primary
  $pp$ interaction vertex. At least one charged particle
  must have $\pt > 1.6\gev$ and be
  inconsistent with originating from a PV.
  A multivariate algorithm~\cite{BBDT} is used for
  the identification of secondary vertices consistent with the decay
  of a \bquark hadron.


 Simulation is required to model the effects of the detector acceptance and the
  imposed selection requirements. Here $pp$ collisions are generated using
\pythia~\cite{Sjostrand:2007gs,*Sjostrand:2006za} with a specific \lhcb
configuration~\cite{LHCb-PROC-2010-056}.  Decays of unstable particles
are described by \evtgen~\cite{Lange:2001uf}, in which final-state
radiation is generated using \photos~\cite{Golonka:2005pn}. The
interaction of the generated particles with the detector, and its response,
are implemented using the \geant
toolkit~\cite{Allison:2006ve, *Agostinelli:2002hh} as described in
Ref.~\cite{LHCb-PROC-2011-006}.

\begin{table}[t]
\begin{center}
\caption{Branching fractions of semileptonic $b$-hadron decays from direct measurements for \Bzb and \Bm mesons, ($\left< B\right>\equiv \left< \Bzb +B^-\right>$), and derived  for $\Bsb$ and \Lb hadrons based on the equality of semileptonic widths and the lifetime ratios~\cite{Tanabashi:2018oca,Bigi:2011gf}.  Corrections to $\Gamma_{\rm SL}$ for \Bsb $(-1.0\pm 0.5)$\% and \Lb $(3.0\pm 1.5)$\% are applied  \cite{Bigi:2011gf}. Correlations in the \Bzb and \Bm branching fraction measurements have been taken into account. See Ref.~\cite{Aaij:2016avz} for more information.}
\vskip 3mm
\label{tab:slpar}
\begin{tabular}{lcccc}
\hline\hline
Particle  &   $\tau$ (ps) & ${\cal{B}}_{\rm SL}$  (\%) & ${\cal{B}_{\rm SL}}$ (\%)  \\
                &  measured     & measured                             & used \\
\hline 
\rule{0pt}{12pt}$\Bzb$ & $1.520\pm 0.004$ & $10.30\pm 0.19$&  $10.30\pm 0.19$  \\
$B^-$ & $1.638\pm 0.004$ & $11.08\pm 0.20$&   $11.08\pm 0.20$  \\
$\left< B\right>$&&$10.70\pm 0.19$ &  $10.70\pm 0.19$  \\
$\Bsb$ & $1.526 \pm 0.015$ &  & $10.24\pm 0.21$    \\
$\Lb$ & $1.470\pm 0.010$ & &  $10.26\pm 0.25$   \\
\hline\hline
\end{tabular}
\vspace*{-4mm}
\end{center}
\end{table}
\begin{table}[b]
\begin{center}
	\caption{Charm-hadron branching fractions for the decay modes used in this analysis. Note, the \Lc branching fraction has been significantly improved since the previous analysis.
	}
	\vskip 3mm
	\label{tab:charmBF}
	\begin{tabular}{lcl}
	\hline\hline
	Decay &  ${\cal{B}}$ (\%)  &~~ ~~~~~Source \\
	\hline
	$D^0\to K^-\pi^+$ 			& $3.93\pm 0.05$  & PDG average \cite{Tanabashi:2018oca} \\
	$D^+\to K^-\pi^+\pi^+$ 		& $9.22\pm 0.17$  & CLEO-c \cite{Bonvicini:2013vxi}\\
	$D_s^+\to K^-K^+\pi^+$ 		& $5.44\pm 0.18$  &PDG average \cite{Tanabashi:2018oca} \\	$\Lc\to pK^-\pi^+$   & $6.23\pm 0.33$    &From Refs.~\cite{Ablikim:2015flg,Zupanc:2013iki} \\
	 \hline\hline
\end{tabular}
\end{center}
\end{table}

Selection criteria are applied to muons and $H_c$ decay particles. The transverse momentum of each hadron must be greater than 0.3\gev, and that of the muon larger than 1.3\gev.  Each track cannot point to any PV, implemented by requiring
$\chisqip >9$ with respect to any PV, where \chisqip\ is defined as the difference in the vertex-fit \chisq of a given PV reconstructed with and
without the track under consideration being included. All final state particles are required to be positively identified using information from the RICH detectors (PID). Particles from $H_c$ decay candidates must have a good fit to a common vertex with $\chi^2$/ndof $<9$, where ndof is the number of degrees of freedom. They must also be well separated from the nearest PV, with the flight distance divided by its uncertainty greater than 5. 

Candidate $b$ hadrons are formed by combining $H_c$ and muon candidates originating from a common vertex with $\chi^2$/ndof $<9$ and an $H_c\mu^-$ invariant mass, $m_{H_c\mu^-}$, in the range 3.0--5.0\gev for $D^0$ and $D^+$, 3.1--5.1\gev for \Ds and 3.3--5.3\gev for \Lc candidates. In addition, we define
$m_{\rm corr} \equiv \sqrt{m_{H_{c}\mu}^{2} + p_{\perp}^{2}} + p_{\perp}$, where $p_{\perp}$ is the magnitude of the combination's momentum component transverse to the $b$-hadron flight direction; we require that $m_{\rm corr}>4.2$ or $4.5\gev$ for \Bsb or \Lb candidates, respectively.
For the $\Ds\to K^+K^-\pi^+$ decay mode, vetoes are employed to remove backgrounds from real $D^+$ or \Lc decays where the particle assignments are incorrect.

Background from prompt $H_c$ production at the PV needs to be considered. We use the natural logarithm of the $H_c$ impact parameter, IP, with respect to the PV in units of mm.
 Requiring  ln(IP/mm)$>-3$ is found to reduce the prompt component to be below 0.1\%,  while preserving 97\% of all signals. This restriction allows us to perform fits only to the $H_c$ candidate mass spectra to find the $b$-hadron decay yields.

\begin{figure}[t]
\vskip -2mm
		\centering
			\includegraphics[scale=0.33]{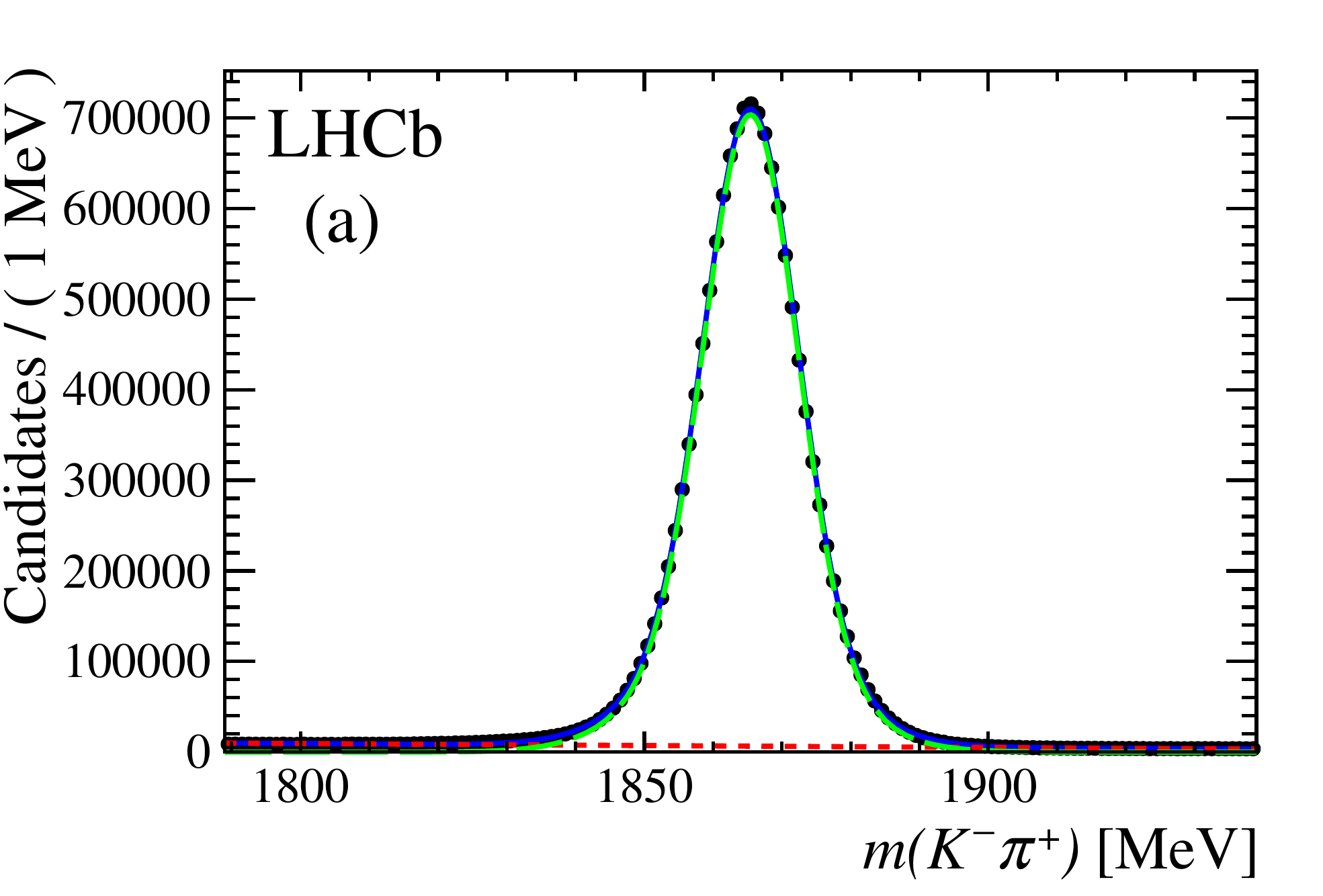}\includegraphics[scale=0.33]{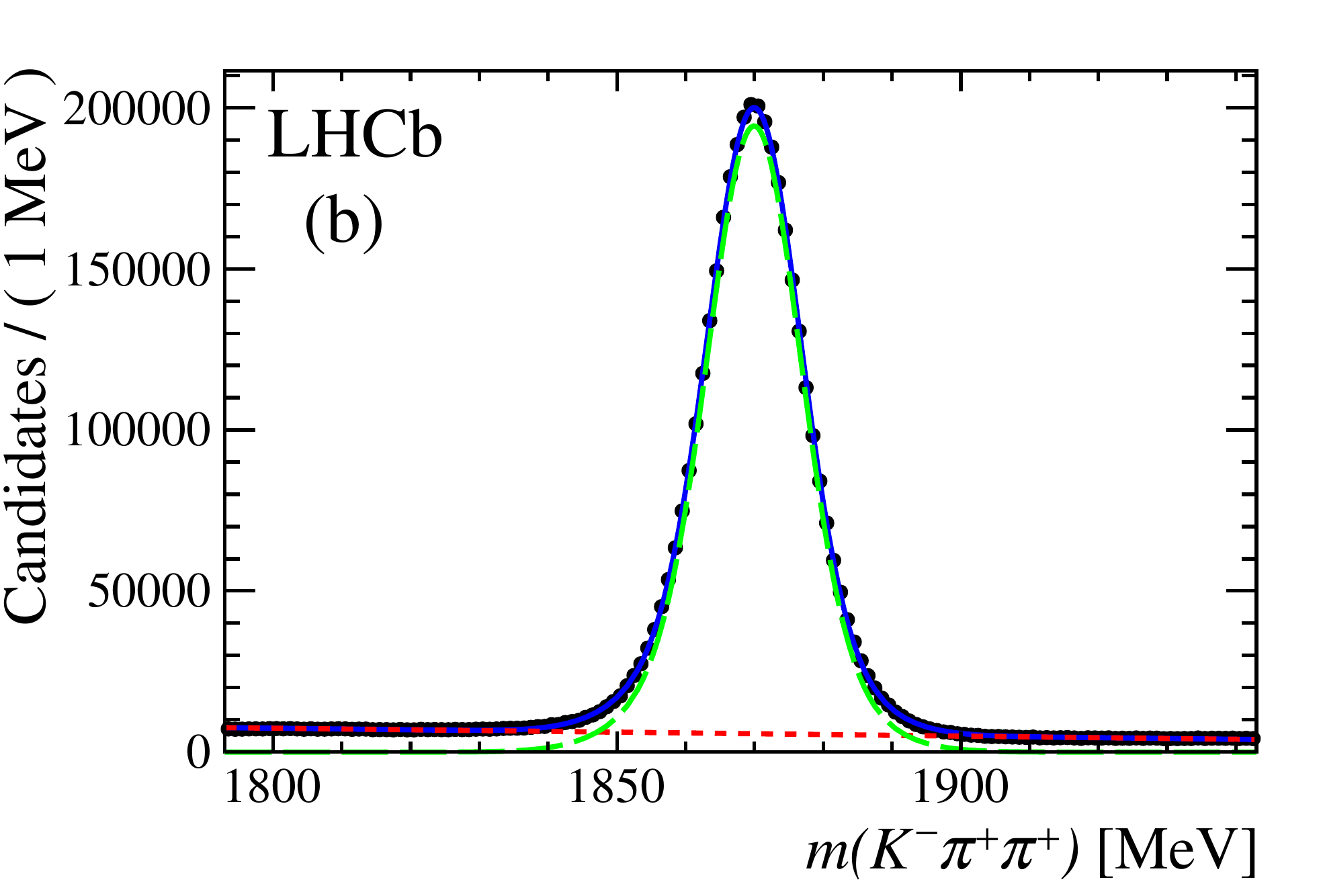}			                   \includegraphics[scale=0.33]{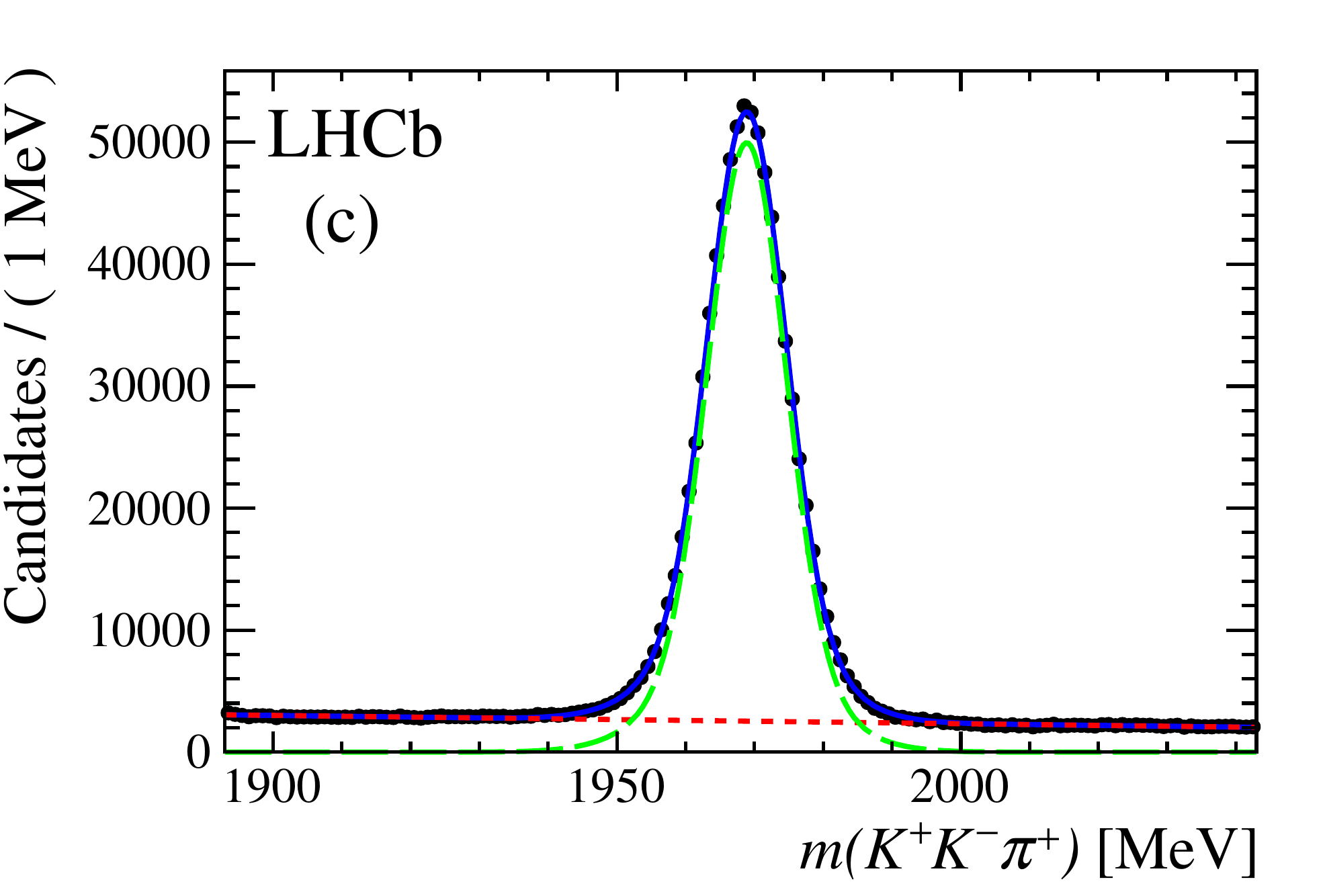}\includegraphics[scale=0.33]{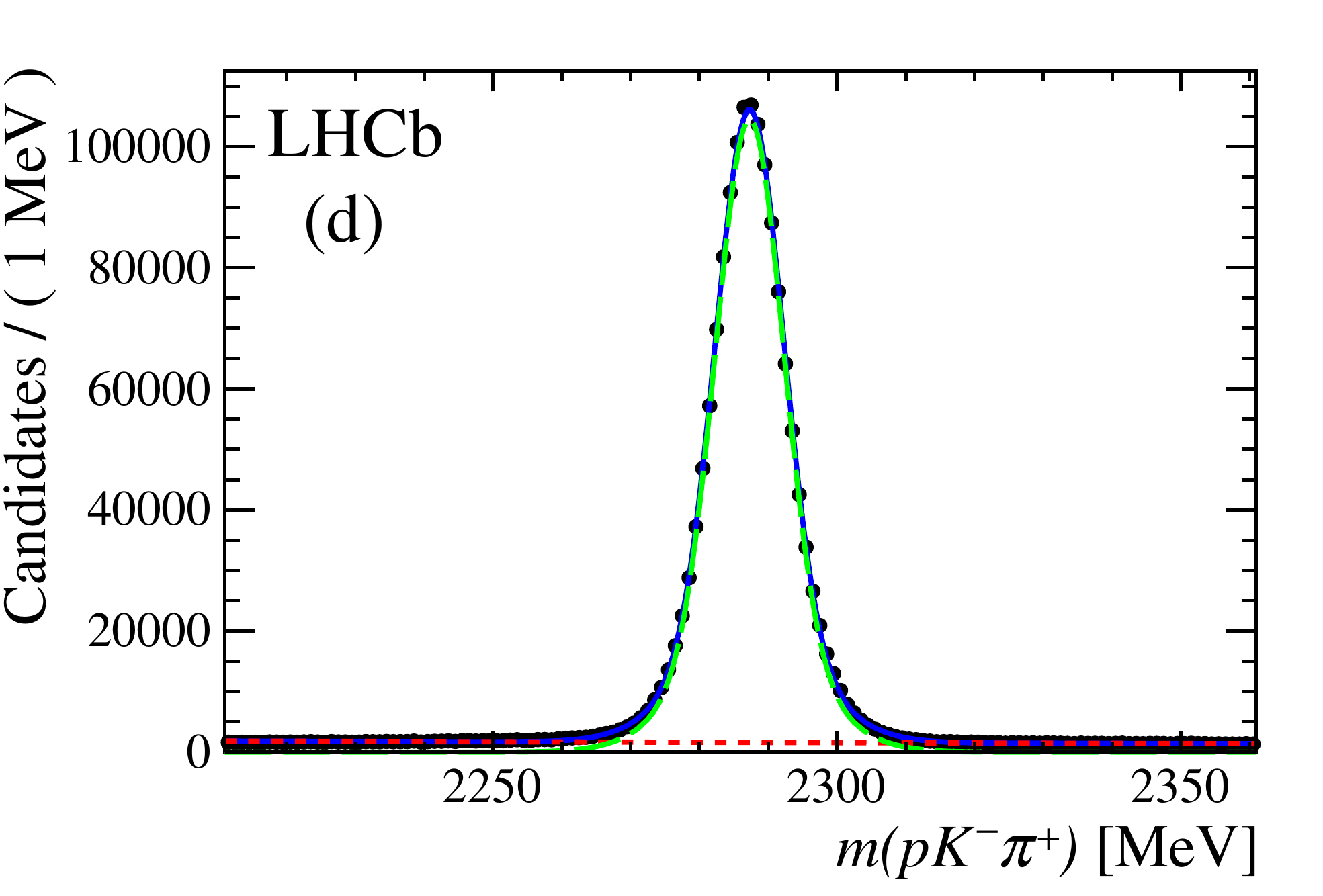}
		\caption{Fit to the mass spectra of the $H_c$ candidates of the selected $H_b$ decays: (a) $D^0$, (b) $D^+$, (c) $D_s^+$ mesons, and (d) the \Lc baryon.  The data are shown as black points with error bars. The signal component is shown as the dashed (green) line and the combinatorial background component is shown as the dashed (red) line. The solid (blue) line shows all components added together.\label{fig:charm-mass}}
\end{figure}

The $H_c$ candidates mass distributions integrated over $\pt(H_b)$ and $\eta$ are shown in Fig.~\ref{fig:charm-mass}. They consist of a prominent peak resulting from signal, and a small contribution due to combinatorial background from random combinations of particles that pass the selection. They are fit with a signal component comprised of two Gaussian functions, and a combinatorial background component modeled as a linear function. The total signal yields for $D^0 X\mu^-\overline{\nu}_{\mu}$, $D^+ X\mu^-\overline{\nu}_{\mu}$, $D_s^+ X\mu^-\overline{\nu}_{\mu}$ and $\Lc\mu^-X\overline{\nu}_{\mu}$ are 13~$\!$775~$\!$000, 4~$\!$282~$\!$700,  845~$\!$300, and 1~$\!$753~$\!$600, respectively.

Background contributions to the $b$-hadron candidates include hadrons faking muons, false combinations of charm hadrons and muons from the two $b$ hadrons in the event, as well as real muons and charm hadrons from $B\to \D\Db X$ decays, where one of the $D$ mesons decays into a muon. All the backgrounds are evaluated in two-dimensional $\eta$ and $\pt$ intervals. The first two backgrounds are evaluated using events where the  $H_c$ is combined with a muon of the wrong-sign ({\it e.g.} $D^0\mu^+$), forbidden in a semileptonic $b$-hadron decay. The wrong-sign backgrounds are $<1$\% for each $H_c$ species. The background from $B\to D\Db X$ decays is determined by simulating a mixture of these decays using their measured branching fractions \cite{Tanabashi:2018oca}. The only decay mode significantly affected is $\Bsb\to \Ds X\mu^-\overline{\nu}_{\mu}$ with contributions varying from 0.1\% for $D^0D_s^- X$  to 1.8\% for $\Ds D_s^- X$ due to the large $\Ds\to\mu^+\nu$ decay rate. The total $B\to D\Db X$ background is $(5.8\pm 0.9)\%$.

The dominant component in \Bsb semileptonic decays is $\Ds X\mu^-\overline{\nu}_{\mu}$, where $X$ contains possible additional hadrons. However, the \Bsb meson also can decay into $D^0 K^+$ or $D^+K^0$ instead of  \Ds, so we must add this component to the \Bsb rate and subtract it from the $f_u+f_d$ fraction. 
Similarly, in \Lb semileptonic decays we find a $D^0pX$ component. The selection criteria for these final states are similar to those for the $D^0 X\mu^-\overline{\nu}_{\mu}$ and $\Lc X\mu^-\overline{\nu}_{\mu}$ final states described above with the addition of a kaon or proton with $\pt>300\mev$ that has been positively identified. A veto is also applied to reject $D^{*+}\to\pi^+D^0$ decays where the pion mimics a kaon or a proton.

These samples contain background, resonant and nonresonant decays. Separation of these components is achieved by using both right-sign ($H_c$ with $\mu^-$) and wrong-sign ($H_c$ with $\mu^+$) candidates.  In addition, the logarithm of the difference between the vertex $\chi^2$ formed by the added hadron track and the $D\mu$ system and the  vertex $\chi^2$  of  the $D\mu$ system, ${\rm ln}(\Delta\chi^{2}_{\rm V})$, provides separation between combinatorial background and nonresonant semileptonic decays.  True resonant and nonresonant $\Bsb\to\Dz\Kp\mu^-\overline{\nu}_{\mu}$ or $\Lb\to\Dz p\mu^- \overline{\nu}_{\mu}$ decays peak in the ${\rm ln}(\Delta\chi^{2}_{\rm V})$ distribution at a value of unity while the background is smooth and rises at higher values as the added track is generally not associated with the $\Dz\mun$ vertex. To distinguish signal from background we define $m(D^{0}h)_C\equiv m(D^0h)-m(D^0)+m(D^0)_{\rm PDG}$, and perform two-dimensional fits to the $m(D^{0}h)_C$ and ${\rm ln}(\Delta\chi^{2}_{\rm V})$ distributions, where $h=K^+(p)$ for right-sign \Bsb  (\Lb) decays. 

\begin{figure}[t]
\vspace{-2mm}
\begin{center}
\includegraphics[width=0.4\textwidth]{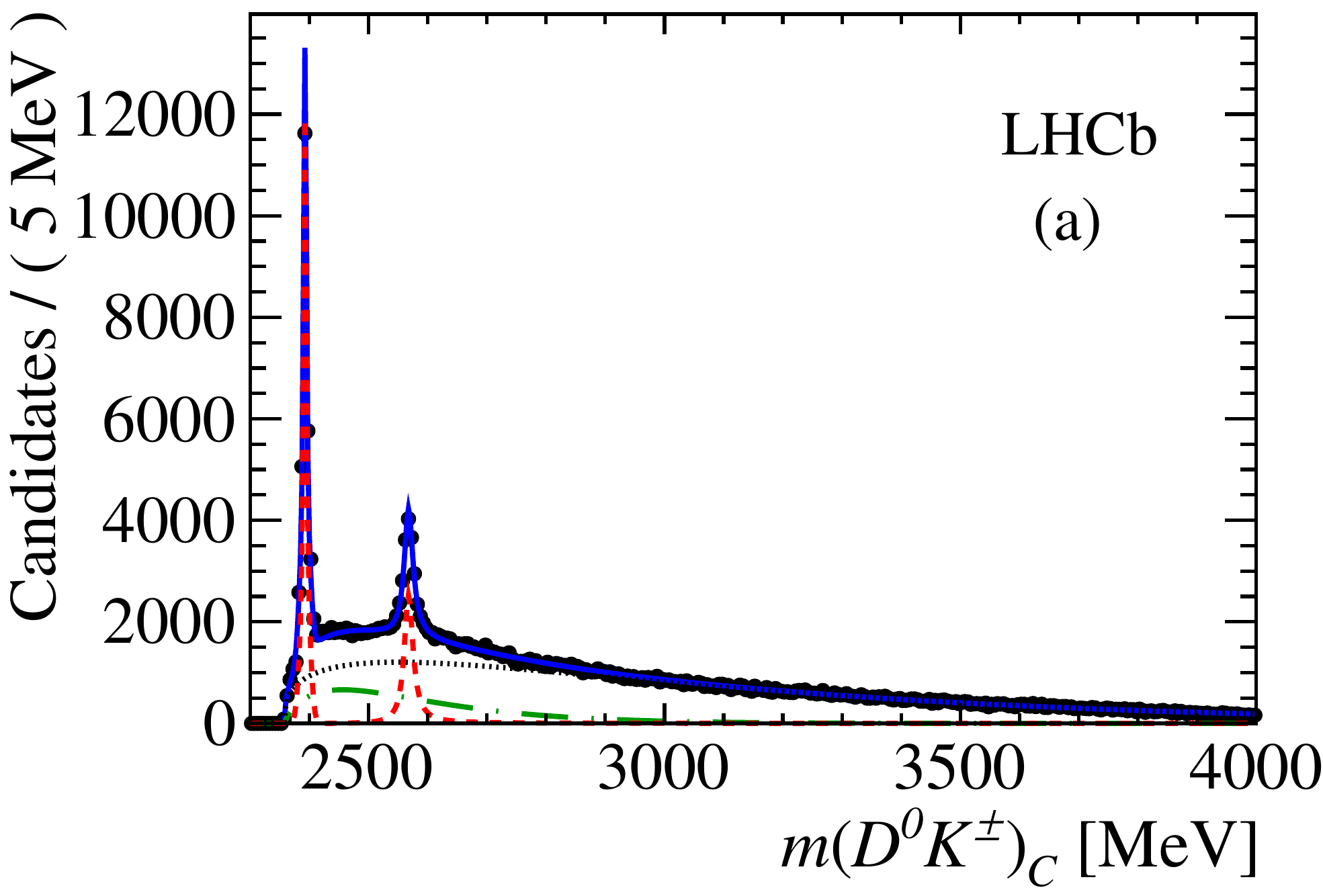}
\includegraphics[width=0.4\textwidth]{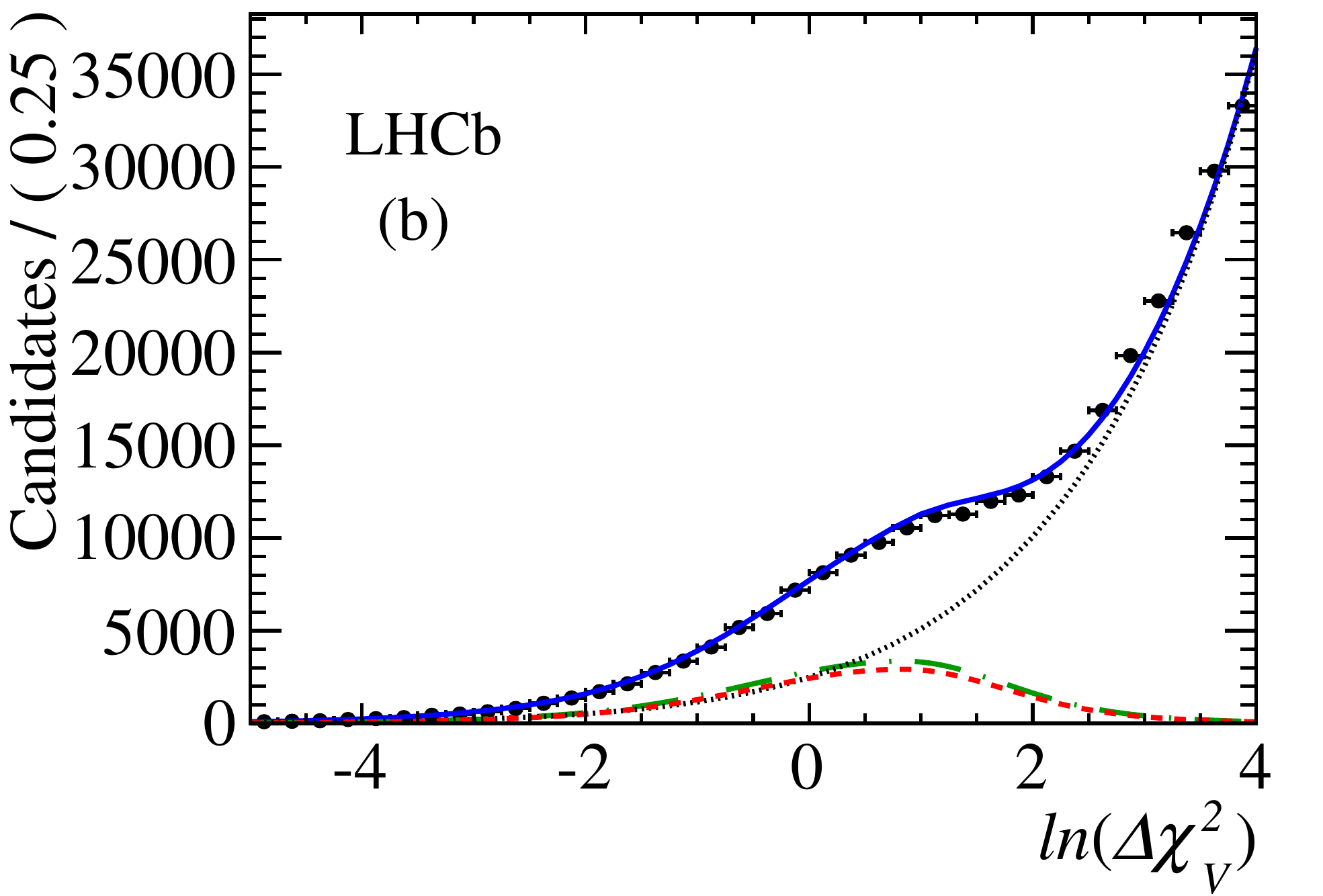}\\
\includegraphics[width=0.4\textwidth]{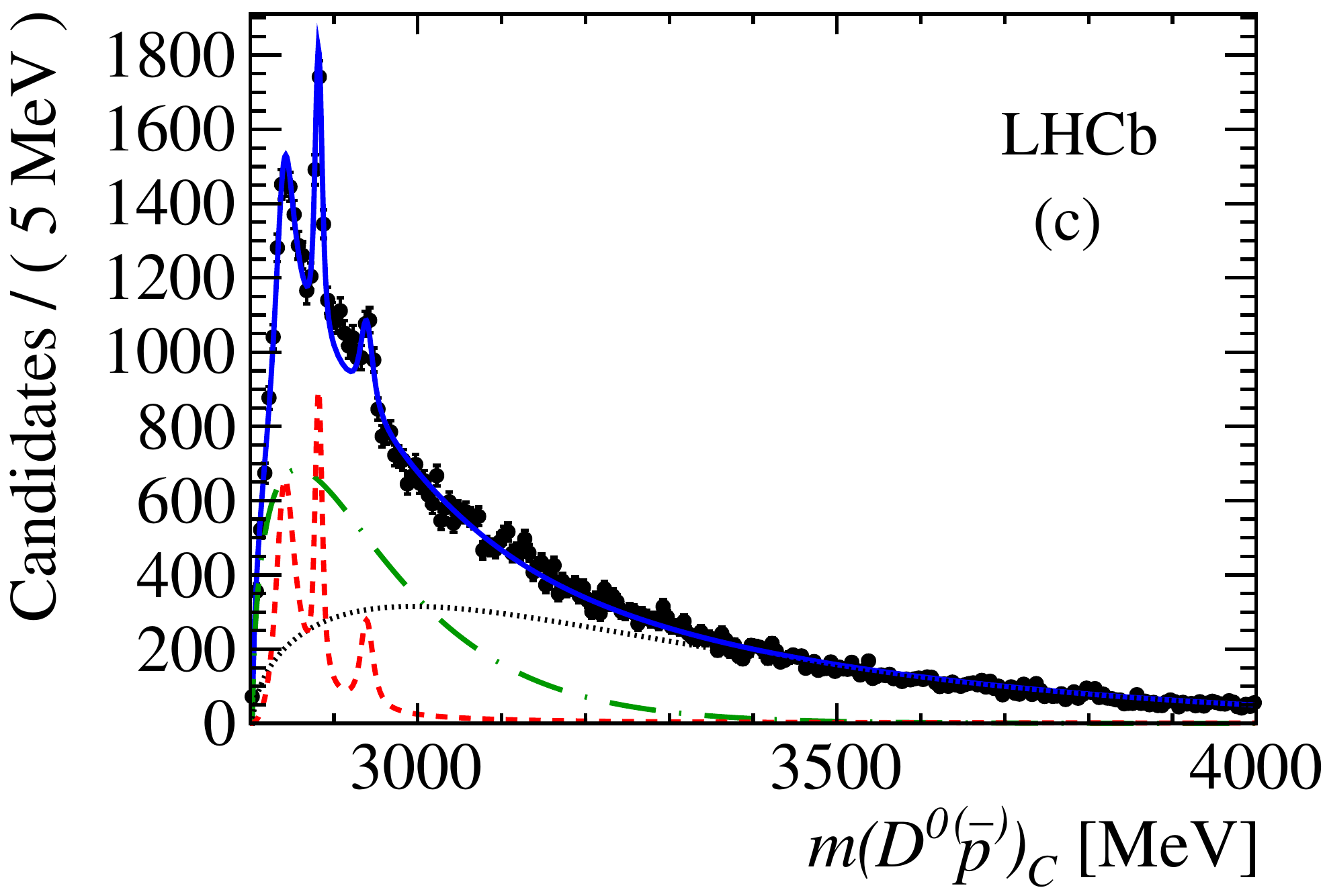}
\includegraphics[width=0.4\textwidth]{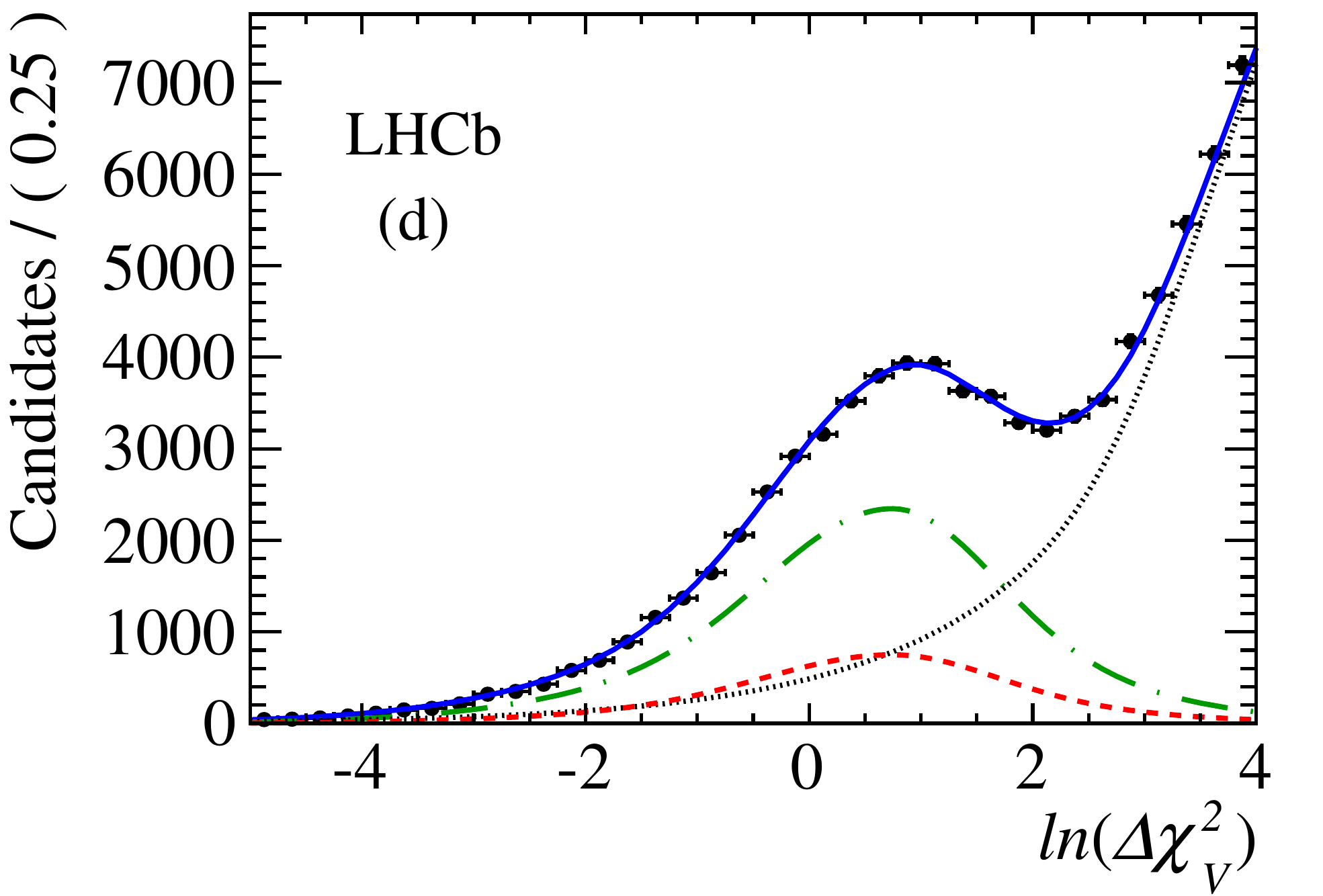}
\caption{ Projections of the two-dimensional fits to the (a)  $m(D^0K^{\pm})_C$ and (c) $m(D^0\porpbar)_C$ mass distributions and (b, d) ${\rm ln}(\Delta\chi^{2}_{\rm V})$ 
for (top) $\Dz K^{\pm} X\mu^- \overline{\nu}_{\mu}$ candidates, and (bottom) for $\Dz \porpbar X\overline{\nu}_{\mu}$ candidates.
The curves show projections of the 2D fit. The  dashed (red) curves show the 
$D_{s1}^{+}$ and $D_{s2}^{*+}$ resonant components in (a) and (b), and $\Lc(2860)$, $\Lc(2880)$ and $\Lc(2940)$
resonant components in (c) and (d). The long-dashed-dotted (green) curves show the nonresonant component, the
dotted (black) curves are the background components, whose shapes are determined from wrong-sign combinations, and the solid (blue) curve shows all components added together.}
\label{fig:full_fits_bs}
\end{center}
\end{figure}
The wrong-sign shapes are used to model the backgrounds.  The resonant structures are modeled with relativistic Breit--Wigner functions convoluted with Gaussians to take into account the experimental resolution, except for the narrow $D_{s1}(2536)^+$ which is modeled with the sum of two Gaussians with a fixed mean. The nonresonant shape for the ${\rm ln}(\Delta\chi^{2}_{\rm V})$ distribution is taken as the same as the resonant one.  Figure~\ref{fig:full_fits_bs} shows the data and result of the fits for \Bsb and \Lb candidates. 

For the \Bsb case, we find $22~\!610\pm210$ $D_{s1}(2536)^+$, $14~\!290\pm260$ $D_{s2}^*(2573)^{+}$, and $38~\!140\pm460$ nonresonant decays, confirming the existence of both the $D_{s1}^+$ \cite{Abazov:2007wg,Aaij:2011ju} and $D_{s2}^{*+}$ \cite{Aaij:2011ju} particles in semileptonic \Bsb decays with substantially more data, and showing the existence of the nonresonant component. To account for the unmeasured $D^+K^0$ channel we take different mixtures of $D^*$ and $D$ final states for the different resonant and nonresonant components. The $D_{s1}^+$ decays dominantly into $D^*$, while the $D_{s2}^{*+}$ decays dominantly into $D$ mesons \cite{Tanabashi:2018oca}. For the nonresonant part we assume equal $D^*$ and $D$ yields. 

In the \Lb case, we find $6120\pm460$ $\Lc(2860)$, $2200\pm200$ $\Lc(2880)$, $1200\pm260$ $\Lc(2940)$, and $29~\!770\pm690$ nonresonant events. The decay rate into $D^0p$ is assumed to be equal to that into $D^+n$ using isospin conservation. All decays with an extra hadron have lower detection efficiencies than the sample without.

Efficiencies for all the samples are determined using data in two-dimensional $\pt$ and $\eta$ bins.
Trigger efficiencies are determined using a sample of \mbox{$B^-\to\jpsi K^-$,} with $\jpsi\to\mup\mun$ decays where only one muon track is positively identified, in conjunction with viewing the effects of combinations of different triggers 
\cite{LHCb-PUB-2014-039}. This sample is also used to determine muon identification efficiencies.
Decays of \jpsi mesons to muons reconstructed using partial information from the tracking system,  {\it e.g.} eliminating the vertex locator information, are also used to determine tracking efficiencies using data and to correct the simulation.
 Finally, the PID efficiencies are evaluated using 
kaons and pions from $D^{*+}\to\pi^+D^0$ decays, with $D^0\to K^-\pi^+$, and protons from $\PLambda\to p \pi^-$ and $\Lc\to p K^-\pi^+$ decays \cite{Aaij:2018vrk}. In the measurement of $b$-hadron fraction ratios many of the efficiencies cancel and we are left with only residual effects to which we assign systematic uncertainties.

The $b$-hadron $\eta$ and \pt, $\pt(H_b)$, must be known because the $b$ fractions can depend on production kinematics.
While $\eta$ can be evaluated directly using the measured primary and secondary $b$ vertices, the value of $\pt(H_b)$  must be determined to account for the missing neutrino plus extra particles. The correction factor $k$ is given by the ratio of the average reconstructed to true $\pt(H_b)$ as a function of $m(H_c\mu^-)$ and is determined using simulation. It varies from 0.75 for $m(H_c\mu^-)$ equals 3\gev to unity at  $m(H_c\mu^-)=m(H_b)$.

\begin{figure}[b]
\vspace{-0.5cm}
		\centering
			\includegraphics[scale=0.6]{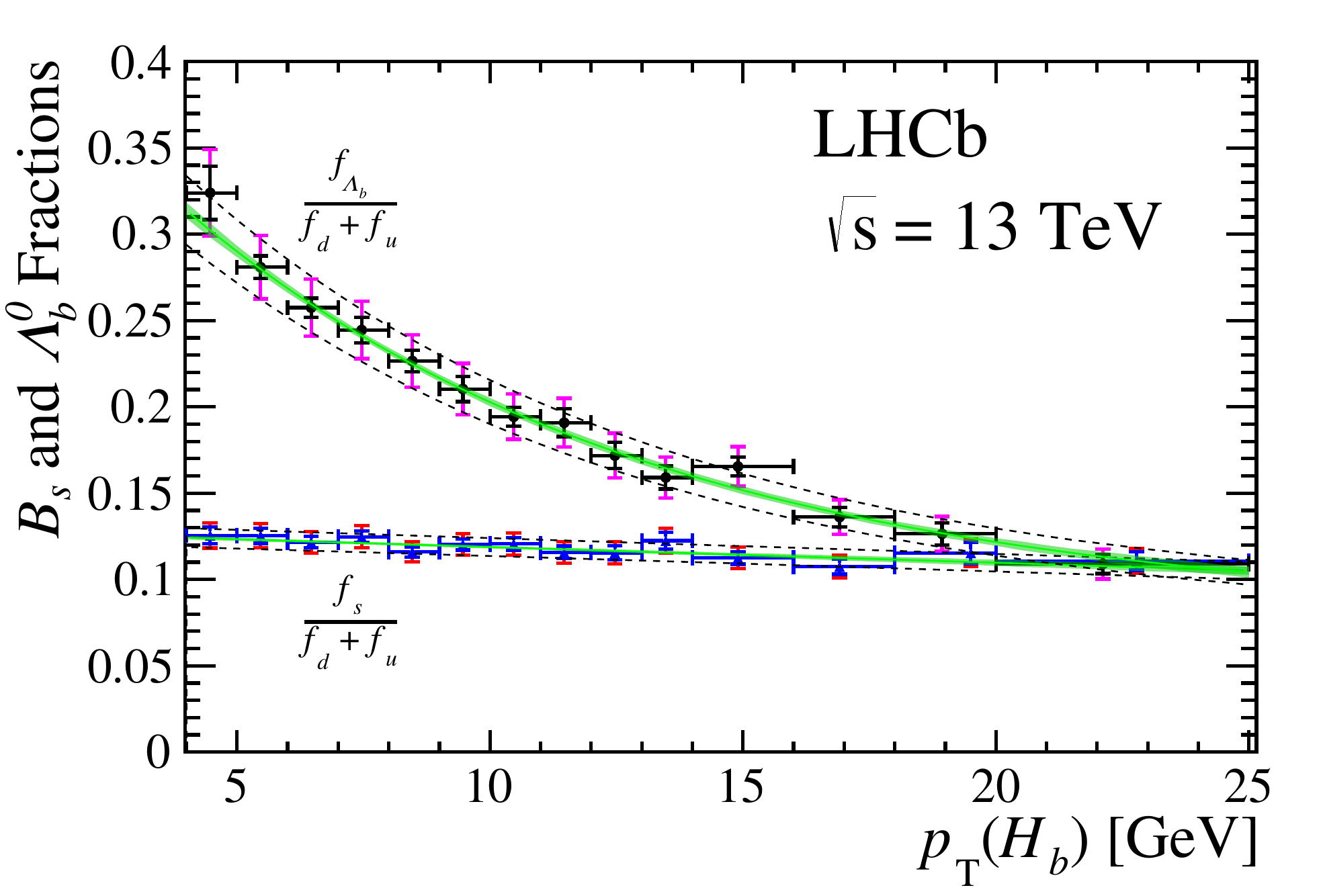}	
			\vspace{-0.25cm}
					\caption{The ratios $f_{s}/(f_{u}+f_{d})$ and $f_{\Lb}/(f_{u}+f_{d})$ in bins of $\pt(H_b)$. The \Bsb data are indicated by solid circles, while the \Lb by triangles. The smaller (black) error bars show the combined bin-by-bin statistical and systematic uncertainties, and the larger (blue) ones show the global systematics added in quadrature. The fits to the data are shown as the solid (green) bands, whose widths represents the $\pm 1 \sigma$ uncertainty limits on the fit shapes, and the dashed (black) lines give the total uncertainty on the fit results including the global scale uncertainty. In the highest two \pt bins the points have been displaced from the center of the bin. \label{fig:fs_fit_k}}
\end{figure}

The distribution of $f_{s}/(f_{u}+f_{d})$ as a function of $\pt(H_b)$ is shown in Fig.~\ref{fig:fs_fit_k}. We perform a linear $\chi^{2}$ fit incorporating a full covariance matrix which takes into account the bin-by-bin correlations introduced from the kaon kinematics, and PID and tracking systematic uncertainties. The factor $A$ in Eq.~\ref{eq:fs} incorporates the global systematic uncertainties described later, which are independent of  $\pt(H_b)$. The resulting function is

\begin{equation}
\label{eq:fs}
\frac{f_{s}}{f_{u}+f_{d}}(\pt) = A\left[p_1 + p_2 \times \left(\pt-\langle\pt\rangle\right)\right],
\end{equation}
where \pt here refers to $\pt(H_b)$, $A=1\pm 0.043$, $p_1=0.119 \pm 0.001$, $p_2=(-0.91 \pm 0.25) \cdot 10^{-3}~\!{\rm GeV}^{-1}$, and  $\langle\pt\rangle=10.1\gev$.
The correlation coefficient between the fit parameters is 0.20. After integrating over $\pt(H_b)$, no $\eta$ dependence is observed (see the Supplemental material).

 We determine an average value for $f_{s}/(f_{u}+f_{d})$ by dividing the yields of $\Bsb$ semileptonic decays by the  sum of $\Bzb$ and $\Bm$ semileptonic yields, which are all efficiency-corrected, between the limits of $\pt(H_b)$ of 4 and 25\gev and $\eta$ of 2 and 5, resulting in
\begin{equation}
\frac{f_{s}}{f_{u}+f_{d}} = 0.122 \pm 0.006,\nonumber
\end{equation}
where the uncertainty contains both statistical and systematic components, with the latter being dominant, and discussed subsequently. The total relative uncertainty is 4.8\%.

Figure~\ref{fig:fs_fit_k} also shows the \Lb fraction as a function of $\pt(H_b)$ demonstrating a large \pt dependence. The distribution in $\eta$ is flat. We perform a similar fit as in the \Bsb fraction case, using
\begin{equation}
\label{eq:fLb}
\frac{f_{\Lb}} {f_{u}+f_{d}}(\pt) =A\left[ p_1 + \exp\left( p_2 +p_3 \times \pt\right)\right],\\
\end{equation}
where \pt here refers to $\pt(H_b)$, $A=1\pm 0.061$, $p_1=( 7.93\pm 1.41)\cdot10^{-2}$, $p_2=-1.022 \pm 0.047$, and \mbox{$p_3= -0.107 \pm 0.002~\!{\rm GeV}^{-1}$}.
The correlation coefficients among the fit parameters are 0.40 $(\rho_{12})$, --0.95 $(\rho_{13})$, and --0.63 $(\rho_{23})$.

The average value for $f_{\Lb} / (f_{u}+f_{d})$ is determined using the same method as in the \Bsb case.  The result is
\begin{equation}
\frac{f_{\Lb}}{f_{u}+f_{d}} = 0.259  \pm 0.018,\nonumber
\end{equation}
where the dominant uncertainty is systematic, and the statistical uncertainty is included. The overall uncertainty is 6.9\%. 

As a systematic check of the analysis method, and a useful measurement to test the knowledge of known semileptonic branching fractions and extrapolations used to saturate the unknown portion of the inclusive hadron spectrum, we measure the ratio of the $D^0 X\mu^-\overline{\nu}_{\mu}$ to $D^+ X\mu^-\overline{\nu}_{\mu}$ corrected yields $f_{+} / f_{0}$. We subtract the small contributions  from \Bsb and \Lb decays, and a very small contribution from $B\to \Ds \Kb \mu^- X$ decays has been taken into account \cite{delAmoSanchez:2010pa}, as in all the fractions measured above.

Assuming $f_{u}$ equals $f_{d}$, Ref.~\cite{Rudolph:2018rzl} estimates the fraction of $D^{+}\mu$ with respect to $D^{0}\mu$ modes in the sum of $\Bm$ and $\Bzb$ decays as $0.387\pm0.012\pm0.026$. The first uncertainty comes from the uncertainties on known measurements. The second uncertainty comes from the different extrapolations from excited $D$ mesons used to saturate the remaining portion of the inclusive rate.

The $f_{+} / f_{0}$  ratio must be independent of $\eta$ and \pt. 
To derive an overall value for $f_{+} / f_{0}$, the $\pt(H_b)$ distribution is fit to a constant.  Only the PID and tracking systematic uncertainties on the second pion in the $D^{+}$ decay need be considered. 
Performing a $\chi^{2}$ fit using the full covariance matrix we find 
$ f_{+} / f_{0}=0.359\pm0.006\pm 0.009$, where the first uncertainty is from bin-by-bin statistical and systematic uncertainties, including correlations, and the second is systematic. The $\chi^2$/ndof is 0.63, in agreement with a flat spectrum. The measurement is consistent with the prediction and places some constraints on the $D^{**}$ content of  semileptonic $B$ decays \cite{Rudolph:2018rzl}.

  The dominant global systematic uncertainties are listed in Table~\ref{tab:Systematics}. Simulation uncertainties are due to the modeling of excited charm states for the $f_s/(f_u+f_d)$ determination and the weighting required for the $f_{\Lb}/(f_u+f_d)$ ratio, due to differences between the simulated and measured \pt spectra. Background uncertainties arise from $D\Db X$ final states with uncertain branching fractions. Cross-feed uncertainties come from errors on efficiency estimates and the assumed $D^*$ to $D$ mixtures.
  Other smaller uncertainties depend on $\pt(H_b)$ and include tracking (0.2--1.8)\%, particle identification (0.4--3.0)\%, trigger (0.3--3.9)\% and $k$-factor (0.2--1.8)\%.

\begin{table}[t]
	\caption{Global systematic uncertainties. The $D^{0}$ and $D^{+}$ branching fraction uncertainties are scaled by the fraction of each decay, $f_0$ and $f_+$ for $f_s/(f_u+f_d)$ and  $f_{\Lb}/(f_u+f_d)$ uncertainties.
	 \label{tab:Systematics}}
	\centering
	\begin{tabular}{lccc}
	 \hline\hline
	Source &  \multicolumn{3}{c}{Value (\%)} \\
	             & $f_s/(f_u+f_d)$ &   $f_{\Lb}/(f_u+f_d)$ & $f_+/f_0$ \\ [0.1cm]
	\hline	
	Simulation 							&  1.7      & 2.4		&--\\
	Backgrounds 							&  0.9       & 0.3		&--\\
	Cross-feeds							&  1.2       & 0.4		& 0.2\\
 	$\mathcal{B}$($D^{0} \rightarrow K^-\pi^+$)	&  1.0 	&1.0		& 1.3  \\
	$\mathcal{B}$($D^{+} \rightarrow K^+\pi^-\pi^-$) & 0.6 	& 0.6		& 1.8\\
	$\mathcal{B}$($D^{+}_{s} \rightarrow K^+K^-\pi^+$) & 3.3 &      --  	&-- \\
	$\mathcal{B}$($\Lc \rightarrow pK^+\pi^-$) 		& --   		& 5.3&-- \\
	Measured lifetime ratio					& 1.2 	& 0.7  	& --\\
	$\Gamma_{SL}$ correction					& 0.5   	& 1.5	 	&-- \\
	\hline
	Total  								& 4.3		& 6.1	        &2.2\\
	\hline\hline
	\end{tabular}
	
\end{table}

In conclusion, we measure the ratios of \Bsb and \Lb production to the sum of \Bm and \Bzb to be $\pt(H_b)$ dependent
(see Eqs.~\ref{eq:fs} and \ref{eq:fLb}).
The averages in the ranges $4<\pt(H_b)<25\gev$, and $2<\eta< 5$ are \mbox{$f_s/(f_u+f_d)=0.122 \pm 0.006$},  and \mbox{$f_{\Lb}/(f_u+f_d)=0.259  \pm 0.018$}, respectively. Using 7\tev data, LHCb determined $f_{s}/(f_u+f_d)=0.1295  \pm 0.0075$ with a $\pt(H_b)$ slope larger than, but consistent with these 13\tev results \cite{Aaij:2013qqa}; no dependence on $\eta$ was observed.  For the $\Lb$ baryon, the fraction ratio is consistent with the 7\tev measurements after taking into account the different $\pt(H_b)$ ranges used \cite{Aaij:2014jyk,Aaij:2015fea,Aaij:2011jp}. We observe no rapidity dependence over a similar $\pt(H_b)$ range as in Ref.~\cite{Aaij:2015fea}.

These results are crucial for determining absolute branching fractions of \Bsb and \Lb hadron decays in LHC experiments. 
We also determine the ratio of $D^0$ to $D^+$ mesons produced in the sum of  $\Bzb$ and \Bm semileptonic decays as \mbox{$f_{+} / f_{0}=0.359\pm0.006\pm 0.009$}.

\section*{Acknowledgements}
%
%
\noindent We express our gratitude to our colleagues in the CERN
accelerator departments for the excellent performance of the LHC. We
thank the technical and administrative staff at the LHCb
institutes.
We acknowledge support from CERN and from the national agencies:
CAPES, CNPq, FAPERJ and FINEP (Brazil); 
MOST and NSFC (China); 
CNRS/IN2P3 (France); 
BMBF, DFG and MPG (Germany); 
INFN (Italy); 
NWO (Netherlands); 
MNiSW and NCN (Poland); 
MEN/IFA (Romania); 
MSHE (Russia); 
MinECo (Spain); 
SNSF and SER (Switzerland); 
NASU (Ukraine); 
STFC (United Kingdom); 
NSF (USA).
We acknowledge the computing resources that are provided by CERN, IN2P3
(France), KIT and DESY (Germany), INFN (Italy), SURF (Netherlands),
PIC (Spain), GridPP (United Kingdom), RRCKI and Yandex
LLC (Russia), CSCS (Switzerland), IFIN-HH (Romania), CBPF (Brazil),
PL-GRID (Poland) and OSC (USA).
We are indebted to the communities behind the multiple open-source
software packages on which we depend.
Individual groups or members have received support from
AvH Foundation (Germany);
EPLANET, Marie Sk\l{}odowska-Curie Actions and ERC (European Union);
ANR, Labex P2IO and OCEVU, and R\'{e}gion Auvergne-Rh\^{o}ne-Alpes (France);
Key Research Program of Frontier Sciences of CAS, CAS PIFI, and the Thousand Talents Program (China);
RFBR, RSF and Yandex LLC (Russia);
GVA, XuntaGal and GENCAT (Spain);
the Royal Society
and the Leverhulme Trust (United Kingdom);
Laboratory Directed Research and Development program of LANL (USA).

\clearpage
\newpage
\section{Supplemental material}
\subsection{\boldmath Relationships between raw $H_c\mu^-X$ measured yields and corrected yields}

The corrected yields for $\Bzb$ or $B^-$ mesons decaying into $D^0\munu X$ or $D^+\munu X$, $n_{\rm corr}$, can be expressed in terms of the measured yields, $n$, as

\begin{align}
\label{eq:dzero}
n_{\rm corr}(B\to \Dz \mun)&=\frac{1}
{{\cal{B}}(\Dz\to \Km\pip)\epsilon(B \to \Dz)}\times \\\nonumber 
&\left[n(\Dz\mu^-)-n(\Dz\Kp\mun)
\frac{\epsilon(\Bsb \to \Dz)}{\epsilon(\Bsb \to \Dz\Kp)}-n(\Dz p\mun)
\frac{\epsilon(\Lb \to \Dz)}{\epsilon(\Lb \to \Dz p)}\right],
\end{align}
where we use the  shorthand $n(D\mun)\equiv n(DX\munu)$.  An analogous abbreviation $\epsilon$ is used for the total trigger and detection efficiencies.  
For example, the ratio \mbox{$\epsilon(\Bsb \to D^0K^+)/\epsilon(\Bsb \to D^0)$} gives the relative efficiency to reconstruct a charged kaon in semimuonic $\Bsb$ decays producing a $\Dz$ meson. The second term in this equation accounts for the $\Dz\mun$ pairs originating from a $\Bsb$ decay, such as $\Bsb\to\Dz\Kp\mun$, while the third term accounts for the $\Dz\mun$  pairs originating from $\Lb$ semileptonic decays. These components are determined from the study of the final states  $\Dz\Kp\mun$ and  $\Dz p\mun$respectively. The branching fraction ${\cal B}(\Dz\to\Km\pip)$ appears because this decay mode is used in this study.
Similarly
\begin{eqnarray}
n_{\rm corr}(B\to \Dp \mun)&=&\frac{1}{\epsilon(B \to \Dp)}\left[
\frac{n(\Dp\mun)}{{\cal{B}}(\Dp\to \Km\pip\pip)} \right. \nonumber \\  
&&\left. -\frac{n(\Dz\Kp\mun)}{{\cal{B}}(\Dz\to \Km\pip)}
\frac{\epsilon(\Bsb \to \Dp)}{\epsilon(\Bsb \to \Dz\Kp)}  \right. \nonumber \\
&&\left.
-\frac{n(\Dz p\mun)}{{\cal{B}}(\Dz\to \Km\pip)}\frac{\epsilon(\Lb \to \Dp)}{\epsilon(\Lb \to \Dz p)}\right].\label{eq:dp}
\end{eqnarray}
Both the $\Dz X\munu$ and the $\Dp X \munu$ final states contain small components of cross-feed from $\Bsb$ decays to $\Dz K^+ X\munu$ and to $D^+ \Kz X\munu$, and from $\Lb$ decays to $\Dz p X\munu$ and  to $D^+ n X\munu$. Here we use isospin symmetry and infer the contributions by $\Dp\mun$ pairs originating from a $\Bsb$ decay, such as $\Bsb\to\Dp\Kz\mun\neumb$ from the $\Dz\Kp\mun$  final states, and the contributions from $\Lb\to\Dp n\mun\neumb$ from the $ \Dz p\mun$ yields.

The number of $\Bsb\to \Ds X\munu$ decays in the final state is given by
\begin{eqnarray}
n_{\rm corr}(\Bsb\to\Ds \mu^-)&=& \frac{n(\Ds\mu^-)}{{\cal B}(\Ds\to K^+K^-\pi^+)\epsilon(\Bsb\to\Ds\mu^-)}  \nonumber \\  \label{eq:bsds}
& &-N(\Bzb+B^-){\cal B}(B\to \Ds \Kb^0)\frac{\epsilon(\Bb\to\Ds \Kb^0\mu^-)}{\epsilon(\Bsb\to\Ds \mu^-)}.
\end{eqnarray}
In addition, the $\Bsb$ meson decays semileptonically into $DKX\munu$, and thus we need to add to Eq.~\ref{eq:bsds} the term
\begin{equation}
n_{\rm corr}(\Bsb\to DK\mu^-)=\kappa \frac{n(\Dz K^+\mu^-)}{{\cal B}(\Dz\to K^-\pi^+)\epsilon(\Bsb\to D^0K^+\mu^-)}, \label{eq:bsdk}
\end{equation}
where $\kappa$ accounts for the unmeasured $\Bsb\to D^+KX\munu$ semileptonic decays. The correction $\kappa$ is evaluated using the known decay modes of the $D_{s1}(2536)^+$ and $D^*_{s2}(2573)^+$  states and assuming that the nonresonant component of the hadronic mass spectrum decays in equal portions into $D$ or $\Dstar$ final states. The last term in Eq.~\ref{eq:bsds} accounts for $\Ds K X\munu$ final states originating from $\Bzb$ or $B^-$ semileptonic decays, and $N(\Bzb+B^-)$ indicates the total number of $\Bzb$ and $B^-$ produced. We derive this correction using the PDG value for the branching fraction ${\cal B}(B^-\to D_s^{(*)+}K^-\mu^-\nu)=(6.1\pm 1.0)\times 10^{-4}$, and assuming the same rate for \Bsb decays using isospin invariance \cite{Tanabashi:2018oca}.

The equation for the ratio $\rbs$ is 
\begin{eqnarray}
\label{eq:frac-final}
\frac{\fs}{\fu+\fd}=\frac{n_{\rm corr}(\Bsb\to D\mu^-)}{n_{\rm corr}(B\to D^0 \mu^-)+n_{\rm corr}(B\to D^+ \mu^-)}
\frac{\tau_{B^-}+\tau_{\Bzb}}{2\tau_{\Bsb}}(1-\xi_{s})  \nonumber \\ 
- \frac{\mathcal{B}(B\rightarrow D_{s}\Kb\mu^-)}{\left<\mathcal{B}_{\rm SL}\right >} \frac{\epsilon(\Bb\rightarrow D_{s}^{+})}{\epsilon(\Bsb\rightarrow D_{s}^{+})} ,
\end{eqnarray}
where $\Bsb\to D\mu$ represents  $\Bsb$ semileptonic decays to a charmed hadron, given by the sum of the contributions shown in Eqs. \ref{eq:bsds} and \ref{eq:bsdk}, and the symbols $\tau _{B_i}$ indicate the $B_i$ hadron lifetimes, that are all well measured~\cite{Tanabashi:2018oca}.  We use the average $\Bsb$ lifetime, $1.526\pm0.015$~ps. This equation assumes equality of the semileptonic widths of all the $b$-hadron species. This is a reliable assumption, as corrections in HQET arise only to order 1/$m_b^2$ and the SU(3) breaking correction is quite small, $(-1.0\pm0.5)\%$ \cite{Bigi:2011gf}. The parameter $\xi_s$ accounts for this small adjustment. The second term is the subtraction of the $B^{-,0}\rightarrow \Ds \Kb X\munu$ component that is reconstructed in the signal sample as described in Eq.~\ref{eq:bsds}. The $\mathcal{B}_{\rm SL}$ term in the denominator is the semileptonic branching fraction of the $\Bsb$ derived using the equality of the semileptonic widths and the measured lifetime of the $\Bsb$, listed in Table~\ref{tab:slpar}.

The $\Lb$ corrected yield is derived in an analogous manner
\begin{equation}
n_{\rm corr}(\Lb\to H_c\mu^-)=\frac{n(\Lc\mun)}{{\cal{B}}(\Lc\to p\Km\pip)\epsilon(\Lb \to \Lc)}
+2\frac{n(\Dz p\mun)}{{\cal{B}}(\Dz\to \Km\pip)\epsilon(\Lb \to \Dz p)},\label{eq:lb}
\end{equation}
where $H_c$ represents a generic charmed hadron. The second term includes the cross-feed channel and the factor of two accounts for the isospin $\Lb \rightarrow D^{+} n\mun$ decay. The $\Lb$ fraction is written as
\begin{equation}
\label{eq:frac-final-lb}
\frac{f_{\Lb}}{\fu+\fd}= \frac{n_{\rm corr}(\Lb\to H_c \mu^-)}{n_{\rm corr}(B\to \Dz \mu^-)+n_{\rm corr}(B\to \Dp \mu^-)}
\frac{\tau_{\Bm}+\tau_{\Bzb}}{2\tau_{\Lb}} (1-\xi_{\Lb}).
\end{equation}
While  we assume near equality of the semileptonic widths of different $b$ hadrons, we apply a small adjustment  $\xi_{\Lb}=(3.0\pm1.5)$\%, to account for the chromomagnetic correction, affecting $b$-flavored mesons but not $b$ baryons~\cite{Bigi:2011gf}. The uncertainty is evaluated with conservative assumptions for all the parameters of the heavy quark expansion.

\newpage
\subsection{\boldmath Table of $b$-fractions versus $\pt(H_b)$}

\begin{table}[h]
	\caption{Values of $f_s/(f_u+f_d)$ and  $f_{\Lb}/(f_u+f_d)$ in each $\pt(H_b)$ bin. The first uncertainty is statistical and incorporates both the uncertainties due to the data sample size and the finite amount of simulated events, while the second is the overall systematic uncertainty, including global and bin-dependent systematic uncertainties.
	 \label{tab:ptB}}
	\centering
	\begin{tabular}{rccc}\hline\hline
	$\pt(H_b)[\!\gev]$ & $f_s/(f_u+f_d)$ &   $f_{\Lb}/(f_u+f_d)$  \\[0.1cm]
	\hline	
4--5   ~~   & $0.125 \pm 0.001 \pm 0.007$ &  $0.324\pm 0.001\pm 0.025$ \\
5--6    ~~  & $0.125\pm 0.001 \pm 0.007$ &  $0.281\pm 0.001\pm 0.018$ \\
6--7    ~~  & $0.122 \pm 0.001 \pm 0.006$ &  $0.257\pm 0.001\pm 0.017$ \\	
7--8    ~~  & $0.125\pm 0.001 \pm 0.006$ &  $0.245\pm 0.001\pm 0.017$ \\
8--9   ~~  & $0.116 \pm 0.001 \pm 0.006 $&  $0.227\pm 0.001\pm 0.015$ \\
9--10 ~~ & $0.120 \pm 0.001 \pm 0.006$&  $0.210\pm 0.001\pm 0.015$ \\
10--11~~~ & $0.121 \pm 0.001 \pm 0.006 $&  $0.194\pm 0.001\pm 0.013$ \\
11--12  ~~~& $0.116 \pm 0.001 \pm 0.006 $&  $0.191\pm 0.001\pm 0.014$ \\
12--13 ~~ & $0.116 \pm 0.001 \pm 0.006 $&  $0.172\pm 0.001\pm 0.013$ \\
13--14 ~~ & $0.122 \pm 0.001 \pm 0.007 $&  $0.159\pm 0.001\pm 0.012$ \\
14--16 ~~ & $0.112 \pm 0.001 \pm 0.006 $&  $0.165\pm 0.001\pm 0.012$ \\
16--18 ~~ & $0.107 \pm 0.001 \pm 0.006 $&  $0.136\pm 0.001\pm 0.010$ \\
18--20 ~~ & $0.115 \pm 0.001 \pm 0.008 $&  $0.126\pm 0.001\pm 0.010$ \\
20--25 ~~ & $0.111 \pm 0.001 \pm 0.007 $&  $0.109\pm 0.001\pm 0.009$ \\		
	\hline\hline
	\end{tabular}
\end{table}

\subsection{\boldmath Fraction ratios as functions of $\eta$}

\begin{figure}[h]
\vspace{-2mm}
		\centering
			\includegraphics[scale=0.35]{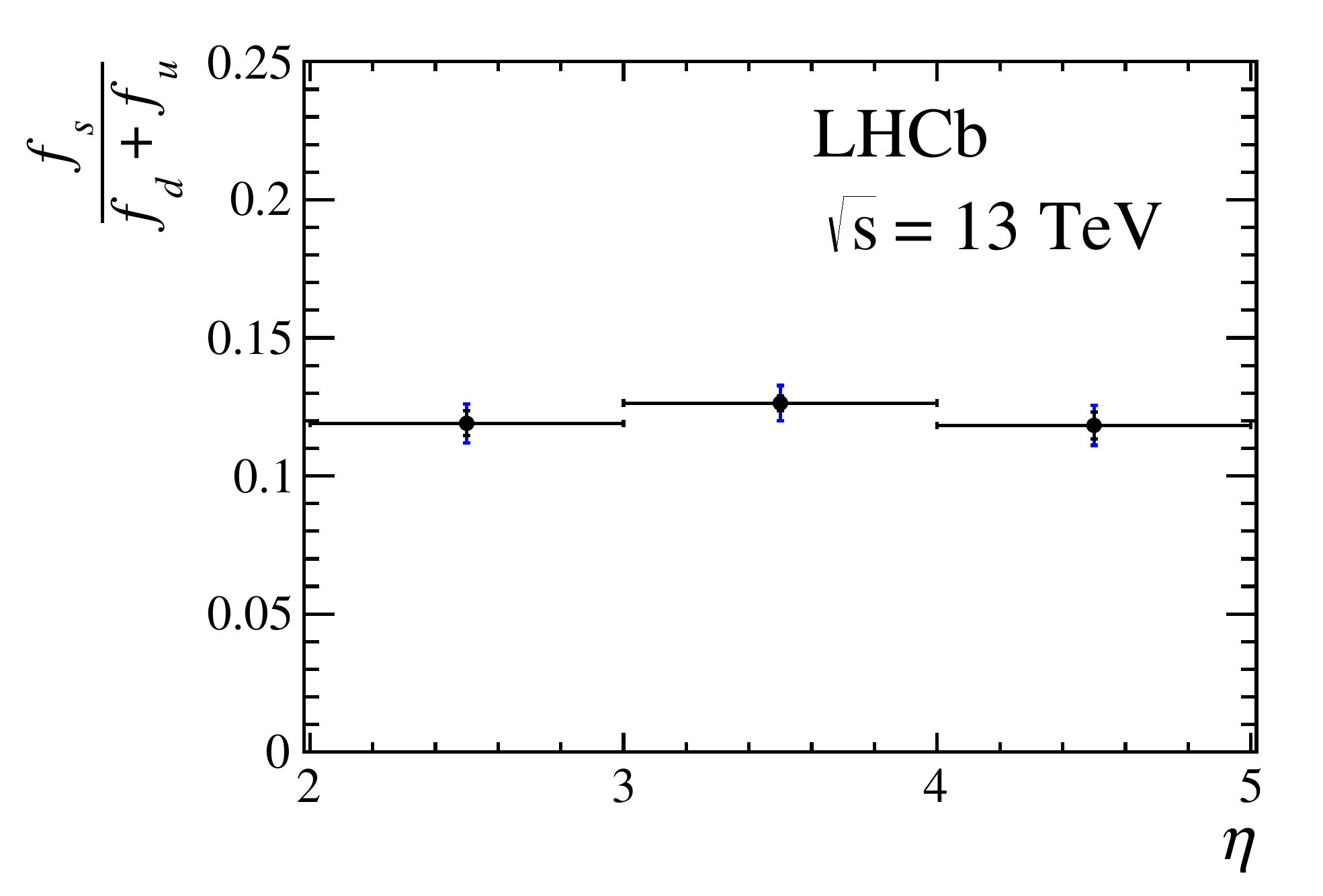}	\includegraphics[scale=0.35]{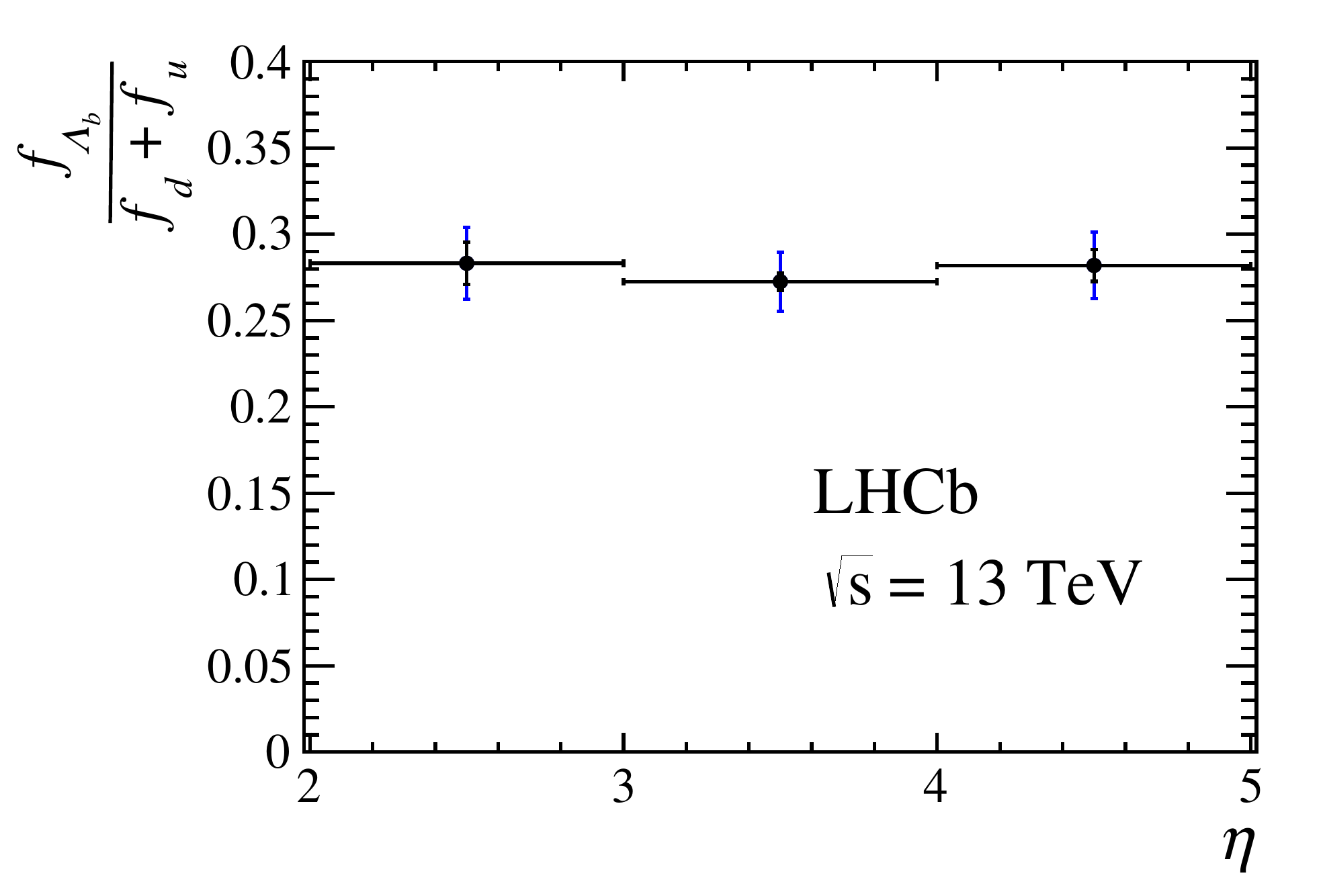}	
		\caption{Measurement of the fraction ratios (a) $f_s/(f_u+f_d)$ and (b) $f_{\Lb}/(f_u+f_d)$ as functions of $\eta$ integrated over \pt.
		 \label{BFraction_eta}}
\end{figure}

Figure~\ref{BFraction_eta} shows measurements of the fraction ratios $f_s/(f_u+f_d)$ and $f_{\Lb}/(f_u+f_d)$ as functions of $\eta$, integrated over \pt. No $\eta$ dependence is visible with the current data sample. 

\subsection{\boldmath Correlation matrices for the fits to $f_s/(f_u+f_d)$ and \mbox{$f_{\Lb}/(f_u+f_d)$}}

Table~\ref{tab:ptfscor} shows the covariance matrix among the different $\pt(H_b)$ bins for $f_s/(f_u+f_d)$, while Table~\ref{tab:ptLbcor} shows the covariance matrix among the different $\pt(H_b)$ bins for $f_{\Lb}/(f_u+f_d)$.
\begin{sidewaystable}
	\caption{Covariance matrix  for $f_s/(f_u+f_d)$ in  $\pt(H_b)$~$\!$[GeV] bins; it accounts for statistical and bin-dependent systematic uncertainties, but not the global systematic uncertainties. \label{tab:ptfscor}}
	\centering
	\footnotesize
	\setlength{\tabcolsep}{3pt}
	\begin{tabular}{rccccccccccccccc}\hline\hline
$\pt(H_b)$ & 4--5&5--6&6--7&7--8&8--9& 9--10&10--11&11--12&12--13&13--14&14--16&16--18&18--20&20--25\\
\hline
4--5&2.30E-05	&3.80E-06&	1.84E-06&	 1.53E-06&	9.28E-07&	1.54E-06&	1.77E-06 &    9.05E-07&      7.47E-07&     6.99E-07&     6.46E-07&     6.84E-07 &5.47E-07	& 5.55E-07\\
5--6& 3.80E-06	&1.96E-05&	2.37E-06&      2.02E-06&	1.23E-06&	2.05E-06&	2.36E-06&	1.22E-06&	1.04E-06&	9.64E-07&	8.94E-07&	9.85E-07 & 8.04E-07	&8.00E-07 \\
6--7& 1.84E-06	&2.37E-06&	1.10E-05&	1.08E-06&	6.71E-07&	1.14E-06&	1.32E-06&	6.94E-07&	5.83E-07&	5.46E-07&	5.18E-07&	5.60E-07& 4.67E-07	&4.89E-07\\
7--8& 1.53E-06&  	2.02E-06	&1.08E-06&	1.08E-05&	6.46E-07&	1.10E-06&	1.27E-06&	6.75E-07&	5.73E-07&	5.43E-07&	5.06E-07&	5.48E-07& 4.80E-07	&4.86E-07\\
8--9 &9.28E-07&	1.23E-06&	6.71E-07	&6.46E-07&	8.73E-06&	7.75E-07&	9.09E-07&	5.07E-07&	4.26E-07&	4.04E-07&	3.79E-07&	4.16E-07&3.77E-07&3.88E-07\\
9--10&1.54E-06&	2.05E-06&	1.14E-06&	1.10E-06&	7.75E-07&	1.09E-05&	1.66E-06&	9.39E-07&	8.03E-07&	7.62E-07&	7.24E-07&	7.77E-07 & 7.01E-07& 7.34E-07\\
10--11&1.77E-06&	2.36E-06&	1.32E-06&	1.27E-06&	9.09E-07&	1.66E-06&	1.33E-05&	1.13E-06&	9.90E-07&	9.33E-07&	9.37E-07&	1.00E-06&9.03E-07	&9.58E-07\\
11--12& 9.05E-07&	1.22E-06&	6.94E-07&	6.75E-07&	5.07E-07&	9.39E-07&	1.13E-06&	1.33E-05&	5.94E-07&	5.64E-07&	5.74E-07&	6.24E-07&5.73E-07	& 6.17E-07\\
12--13& 7.47E-07&	1.04E-06&	5.83E-07&	5.73E-07&	4.26E-07&	8.03E-07&	9.90E-07&	5.94E-07&	1.60E-05&	5.42E-07&	5.62E-07&	6.33E-07&5.73E-07	&6.08E-07\\
13--14& 6.99E-07&	9.64E-07&	5.46E-07&	5.43E-07&	4.04E-07&	7.62E-07&	9.33E-07&	5.64E-07&	5.42E-07&	2.31E-05&	5.31E-07&	6.33E-07&5.98E-07&6.13E-07\\
14--16& 6.46E-07&	8.94E-07&	5.18E-07&	5.06E-07&	3.79E-07&	7.24E-07&	9.37E-07&	5.74E-07&	5.62E-07&	5.31E-07&	1.32E-05&	7.11E-07&6.66E-07	&7.27E-07\\
16--18& 6.84E-07&	9.85E-07&	5.60E-07&	5.48E-07&	4.16E-07&	7.77E-07&	1.00E-06&	6.24E-07&	6.33E-07&	6.33E-07&	7.11E-07&	1.96E-05& 8.08E-07	&9.33E-07\\
18--20& 5.47E-07&	8.04E-07&	4.67E-07&	4.80E-07&	3.77E-07&	7.01E-07&	9.03E-07&	5.73E-07&	5.73E-07&	5.98E-07&	6.66E-07&	8.08E-07&3.71E-05	&9.58E-07\\
20--25&  5.55E-07&	8.00E-07&	4.89E-07&	4.86E-07&	3.88E-07&	7.34E-07&	9.58E-07&	6.17E-07&	6.08E-07&	6.13E-07&	7.27E-07&	9.33E-07 &9.58E-07	&2.93E-05\\

	\hline\hline
	\end{tabular}
\end{sidewaystable}

\begin{sidewaystable}
	\caption{Covariance matrix  for $f_{\Lb}/(f_u+f_d)$ in $\pt(H_b)$~$\!$[GeV] bins; it accounts for statistical and bin-dependent systematic uncertainties, but not the global systematic uncertainties.   \label{tab:ptLbcor}}
	\centering
	\footnotesize
	
	\setlength{\tabcolsep}{3pt}
	\begin{tabular}{rccccccccccccccc}\hline\hline
	$\pt(H_b)$ & 4--5&5--6&6--7&7--8&8--9& 9--10&10--11&11--12&12--13&13--14&14--16&16--18&18--20&20--25\\
	\hline	
		4--5&	2.40E-4	&3.21E-5	&2.08E-5	&5.03E-5	&3.36E-5	&4.21E-5	&1.60E-5	&3.50E-5	&2.47E-5	&1.30E-5	&6.20E-6&	3.10E-6	&2.60E-6	&3.46E-6\\
5--6	&3.21E-5	&4.34E-5	&5.05E-6	&1.30E-5	&8.90E-6	&1.12E-5	&4.47E-6	&9.65E-6	&7.12E-6	&3.73E-6	&1.80E-6&	9.11E-7	&7.82E-7	&1.12E-6\\
6--7	&2.08E-5&	5.05E-6	&2.99E-5	&9.39E-6	&6.55E-6	&8.41E-6	&3.35E-6	&7.43E-6	&5.44E-6	&2.86E-6	&1.43E-6&	7.04E-7	&6.25E-7	&8.96E-7\\
7--8	&5.03E-5	& 1.30E-5	&9.39E-6	&5.32E-5	&1.97E-5	&2.56E-5	&1.05E-5	&2.26E-5	&1.72E-5	    & 9.14E-6	&4.53E-6&	2.33E-6	&2.04E-6	&3.01E-6\\
8--9	&3.36E-5&	8.90E-6	&6.55E-6	&1.97E-5	&3.96E-5	&2.10E-5	&8.88E-6	&1.95E-5	&1.47E-5&	7.91E-6	&3.91E-6&	2.03E-6	&1.75E-6	&2.75E-6\\
9--10	&4.21E-5&	1.12E-5	&8.41E-6	&2.56E-5	&2.10E-5	&5.44E-5	&1.27E-5	&2.81E-5	&2.14E-5&	1.21E-5	&5.88E-6&	3.11E-6	&2.69E-6	&4.44E-6\\
10--11	&1.60E-5&	4.47E-6	&3.35E-6	&1.05E-5	&8.88E-6	&1.27E-5	&3.03E-5	&1.33E-5	&1.10E-5&	6.15E-6	&3.07E-6&	1.67E-6	&1.43E-6	&2.48E-6\\
11--12	&3.50E-5&	9.65E-6	&7.43E-6	&2.26E-5	&1.95E-5	&2.81E-5	&1.33E-5	&6.41E-5	&2.54E-5&	1.42E-5	&7.23E-6&	3.98E-6	&3.44E-6	&5.86E-6\\
12--13	&2.47E-5&	7.12E-6	&5.44E-6	&1.72E-5	&1.47E-5	&2.14E-5	&1.10E-5	&2.54E-5	&5.71E-5&	1.36E-5	&6.78E-6&	3.85E-6	&3.41E-6	&6.10E-6\\
13--14	&1.30E-5&	3.73E-6	&2.86E-6	&9.14E-6	&7.90E-6	&1.21E-5	&6.15E-6	&1.42E-5	&1.36E-5&	4.37E-5	&4.23E-6&	2.47E-6	&7.22E-6	&3.88E-6\\
14--16 &6.20E-6&	1.80E-6	&1.43E-6	&4.53E-6	&3.91E-6	&5.88E-6	&3.07E-6	&7.23E-6	&6.78E-6&	4.23E-6	&3.01E-5&	1.35E-6	&1.21E-6	&2.24E-6\\
16--18	&3.10E-6&	9.11E-7	&7.04E-7	&2.33E-6	&2.03E-6	&3.11E-6	&1.67E-6	&3.98E-6	&3.85E-6&	2.47E-6	&1.35E-6&	3.17E-5	&8.03E-7	&1.54E-6\\
18--20	&2.61E-6&	7.82E-7	&6.25E-7	&2.04E-6	&1.75E-6	&2.67E-6	&1.43E-6	&3.44E-6	&3.41E-6&	2.22E-6	&1.21E-6&	8.03E-7	&4.21E-5	&1.62E-6\\
20--25	&3.46E-6	&1.12E-6	&8.97E-7	&3.01E-6	&2.75E-6	&4.44E-6	&2.48E-6	&5.83E-6	&6.10E-6&	3.88E-6	&2.24E-6&	1.54E-6	&1.62E-6	&3.09E-5\\

	\hline\hline	

	\end{tabular}
\end{sidewaystable}


\newpage
\addcontentsline{toc}{section}{References}
 \ifx\mcitethebibliography\mciteundefinedmacro
\PackageError{LHCb.bst}{mciteplus.sty has not been loaded}
{This bibstyle requires the use of the mciteplus package.}\fi
\providecommand{\href}[2]{#2}

\newpage
\centerline
{\large\bf LHCb Collaboration}
\begin
{flushleft}
\small
R.~Aaij$^{29}$,
C.~Abell{\'a}n~Beteta$^{46}$,
B.~Adeva$^{43}$,
M.~Adinolfi$^{50}$,
C.A.~Aidala$^{77}$,
Z.~Ajaltouni$^{7}$,
S.~Akar$^{61}$,
P.~Albicocco$^{20}$,
J.~Albrecht$^{12}$,
F.~Alessio$^{44}$
% ...

{\footnotesize \it

% ...
$ ^{7}$Universit{\'e} Clermont Auvergne, CNRS/IN2P3, LPC, Clermont-Ferrand, France\\
% ...
$ ^{12}$Fakult{\"a}t Physik, Technische Universit{\"a}t Dortmund, Dortmund, Germany\\
% ...
$ ^{20}$INFN Laboratori Nazionali di Frascati, Frascati, Italy\\
% ...
$ ^{29}$Nikhef National Institute for Subatomic Physics, Amsterdam, Netherlands\\
% ...
$ ^{43}$Instituto Galego de F{\'\i}sica de Altas Enerx{\'\i}as (IGFAE), Universidade de Santiago de Compostela, Santiago de Compostela, Spain\\
$ ^{44}$European Organization for Nuclear Research (CERN), Geneva, Switzerland\\
% ...
$ ^{46}$Physik-Institut, Universit{\"a}t Z{\"u}rich, Z{\"u}rich, Switzerland\\
% ...
$ ^{50}$H.H. Wills Physics Laboratory, University of Bristol, Bristol, United Kingdom\\
% ...
$ ^{61}$University of Cincinnati, Cincinnati, OH, United States\\
% ...
$ ^{63}$Syracuse University, Syracuse, NY, United States\\
% ...
$ ^{77}$University of Michigan, Ann Arbor, United States, associated to $^{63}$\\
% ...
}
\end{flushleft}

\end{document}